%% file: SPMTreadout.tex
\documentclass{article}
\usepackage[utf8]{inputenc}
\usepackage{lscape,epsfig,graphicx,subcaption}
\usepackage[section]{placeins}
\usepackage{makecell}
\usepackage{multirow}
\usepackage{siunitx,amssymb}
\usepackage{authblk}
\usepackage{color}
\usepackage{url}
\usepackage{overpic}
\usepackage {lineno}
\usepackage{textcomp} 
\usepackage{float}

 \newenvironment{keyword}
{\noindent\textbf{Keywords:}}
{}
 \RequirePackage[colorlinks,citecolor=blue,urlcolor=blue,linkcolor=blue]{hyperref}

\title{Embedded underwater front-end electronics for the 3-inch photomultipliers in the JUNO experiment}

\input{Authors_2025-10-22}

\begin{document}

\maketitle


\begin{abstract}
The Jiangmen Underground Neutrino Observatory (JUNO) is a 20-kton liquid scintillator-based, low-radioactivity, multi-purpose neutrino detector located 693 meters (1800 m.w.e.) underground in the Guangdong province, China. To detect scintillation light produced in the target, the detector is equipped with 17,612 20-inch photomultipliers (PMTs), forming the Large PMT system (LPMT).  In addition, 25,600 3-inch photomultipliers (the Small Photomultiplier System or SPMT) are deployed in the gaps between the LPMTs. 
This paper presents the design and performance of the underwater front-end electronics developed for the SPMT system. It details the individual electronics boards and their key components, the inter-board interfaces, the system-level design, and the firmware architecture that supports data acquisition and control. It also outlines mechanical and thermal integration, board validation procedures, and system performance metrics. The readout chain includes digitization of 128 PMT channels per unit, synchronized time-stamping, charge measurement, event packaging, and bandwidth management. Comprehensive validation confirms the system’s readiness to meet JUNO’s stringent physics goals. The underwater electronics achieve noise levels as low as 0.04 photoelectrons with minimal crosstalk (below 0.4\%) and a bandwidth of 57 MB/s, ensuring reliable single photo-electron detection and operation under high-rate conditions. The SPMT system has now been fully integrated and installed in JUNO. Its commissioning and physics performance will be reported in a future publication.
\end{abstract}


\section{Introduction}

The Jiangmen Underground Neutrino Observatory (JUNO) is a multipurpose neutrino experiment located in southern China, 683 meters underground and will house a 20-kiloton liquid scintillator detector. The main scientific goal of JUNO is to determine the neutrino mass ordering (NMO) by studying the oscillation pattern of reactor antineutrinos at a baseline of $\sim$53 km \cite{JUNO:2015zny}\cite{JUNO:2021vlw} \cite{JUNO:2024jaw} from the Yangjiang and Taishan nuclear power plants. The experiment also aims to achieve sub-percent precision on the oscillation parameters ($\sin^{2}\theta_{12}$, $\Delta m^{2}_{21}$ and $\Delta m^{2}_{31}$) \cite{JUNO:2022mxj} by analyzing a large statistics of antineutrino interactions. In addition to reactor neutrinos, JUNO is sensitive to neutrinos from other origins such as Core Collapse Supernova (CCSN) \cite{JUNO:2023dnp}, Diffuse Supernova Neutrino Background (DSNB) \cite{JUNO:2022lpc}, the Sun \cite{JUNO:2020hqc}\cite{JUNO:2023zty}, the Atmosphere \cite{JUNO:2021tll} and the Earth \cite{JUNO:2025sfc}. 

The JUNO Central Detector (CD) consists of 20-kton of liquid scintillator (LS) contained in a 35.4 m diameter acrylic vessel and supported by a stainless steel structure \cite{JUNO:2023ete}. The light produced by the neutrino interactions in the LS is detected by two photodetection systems: 17,612 20-inch PMTs (LPMT system) and 25,600  3-inch PMTs (SPMT system), respectively. Thanks to the optimization of the LS composition, improving light yield and transparency~\cite{JUNO:2020bcl}, combined with the $\sim$78\% PMT photocoverage and $\sim$30\% detection efficiency, the detector achieves an unprecedented energy resolution of $3\%/\sqrt{E{\rm (MeV)}}$ is achieved \cite{JUNO:2024fdc}. Furthermore, a comprehensive calibration strategy will ensure control of the energy scale to better than 1\%\cite{JUNO:2020xtj}.
The CD is submerged in a 35-kton water Cherenkov detector, instrumented with 2,400 20-inch PMTs. The water acts as a protection against the natural radioactivity from the surrounding rocks. The water Cherenkov detector, combined with the Top Tracker placed above it \cite{JUNO:2023cbw}, is designed to track nearly all muons crossing the detector and veto their associated neutrons and spallation products. The detector is installed in a laboratory 693.35 meters underground (1800 m.w.e.) and began data collection in summer 2025.

The LPMT system, with its large photocoverage of 75\%, drives the energy resolution of the JUNO detector, enabling the detection of approximately 1,600 photoelectrons (PE) per MeV~\cite{JUNO:2024fdc}. By adding the 3-inch PMTs in the gaps between the 20-inch PMTs, the photocoverage is slightly increased to 78\%, with about 40 PE detected per MeV. More importantly, the SPMT system provides a complementary and independent readout to the LPMT system. The 3-inch PMTs primarily operate in single photoelectron (PE) counting mode for almost all events within the Central Detector in the [1--10 MeV] range, owing to their small size. This operation mode makes them unaffected by potential instrumental non-linearities at high illumination levels, unlike the LPMTs. Consequently, the SPMT system can help disentangle systematic uncertainties from the LPMTs and improve the energy scale of the reactor antineutrino energy spectrum~\cite{JUNO:2020xtj, Cabrera_2024}. Finally, with excellent Transit Time Spread (TTS) performances (1.6 ns) \cite{Cao:2021wrq}, they will improve the reconstruction of muon tracks, and among others help for the detection of supernova neutrinos or proton decay \cite{JUNO:2022qgr}. Working in the single photon regime, the signal from the 3-inch PMTs is digitized and the readout electronics only deliver information about charge and photons arrival time (Q/T pairs). Not recording the waveforms, the electronic readout was designed in a more compact way than the LPMT system, reducing the required number of electronic boards and cables, leading to 128 channels handled by one set of electronics. The purpose of this paper is to give an overview of the architecture and performances of the JUNO SPMT frond-end electronics.

The paper is organized as follows: Section~2 recaps the guidelines driving the SPMT electronics requirements; Section~3 describes the SPMT front-end readout electronics architecture, emphasizing the role of each board; Section~4 presents validation and performance results of the overall front-end electronics with 128 3-inch PMTs; conclusions are drawn in Section~5.

\section{The Small PMT system electronics requirements}

\begin{figure}[ht!]
    \centering
    \includegraphics[width=0.8\textwidth]{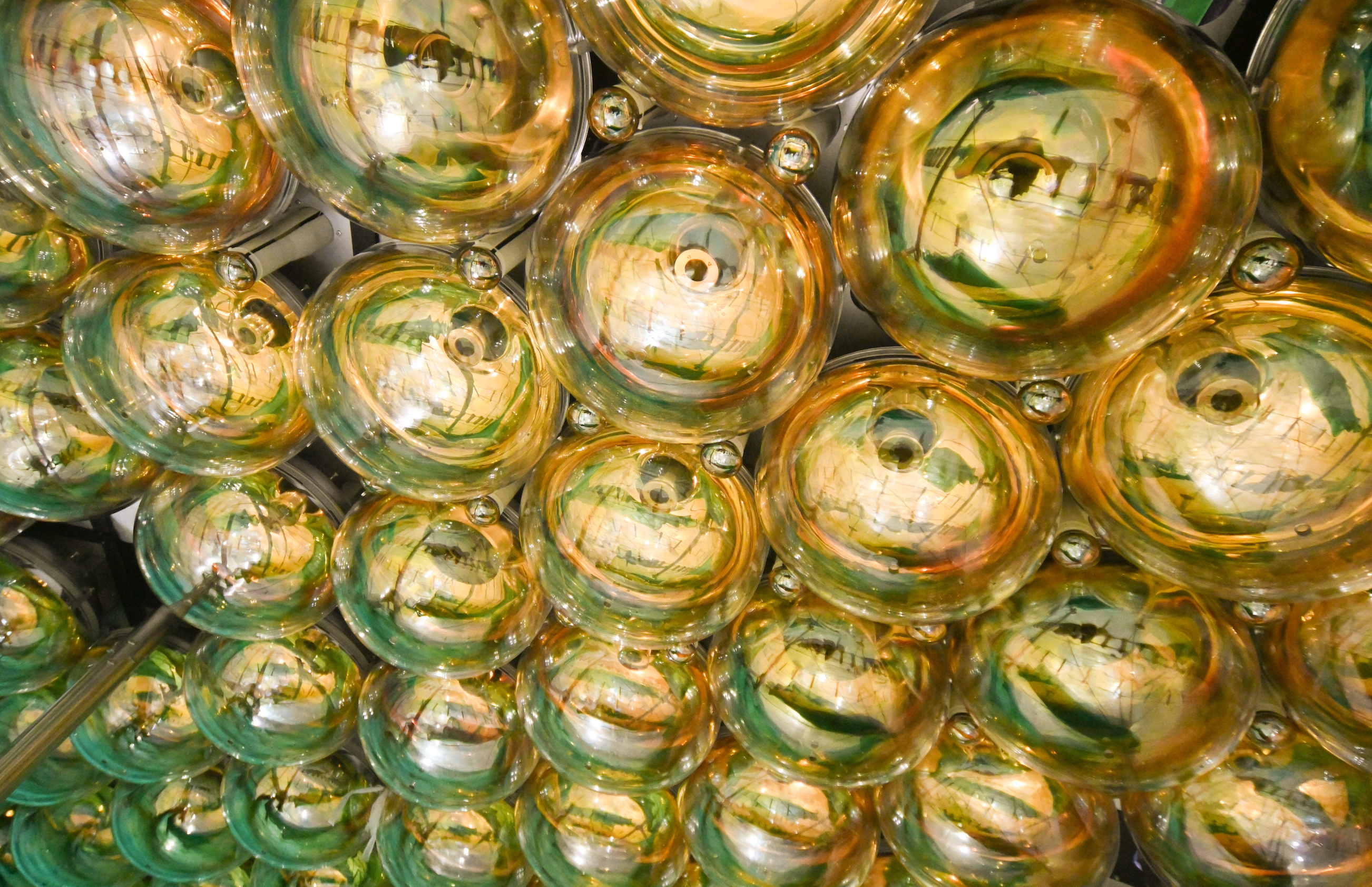}
    \caption[SPMT interlaced]{\label{SPMT interlaced} The 3" PMTs are positioned in the gaps of the 20" LPMTs.}
\end{figure}

The small PMT system of the JUNO experiment consists of 25,600 3-inch PMTs (SPMTs) installed in the gaps between the 20-inch PMTs (LPMTs). The 3-inch size is the maximum that fits between the LPMTs, as shown in Figure~\ref{SPMT interlaced}.

\begin{figure}[ht!]
    \centering
    \includegraphics[width=0.9\textwidth]{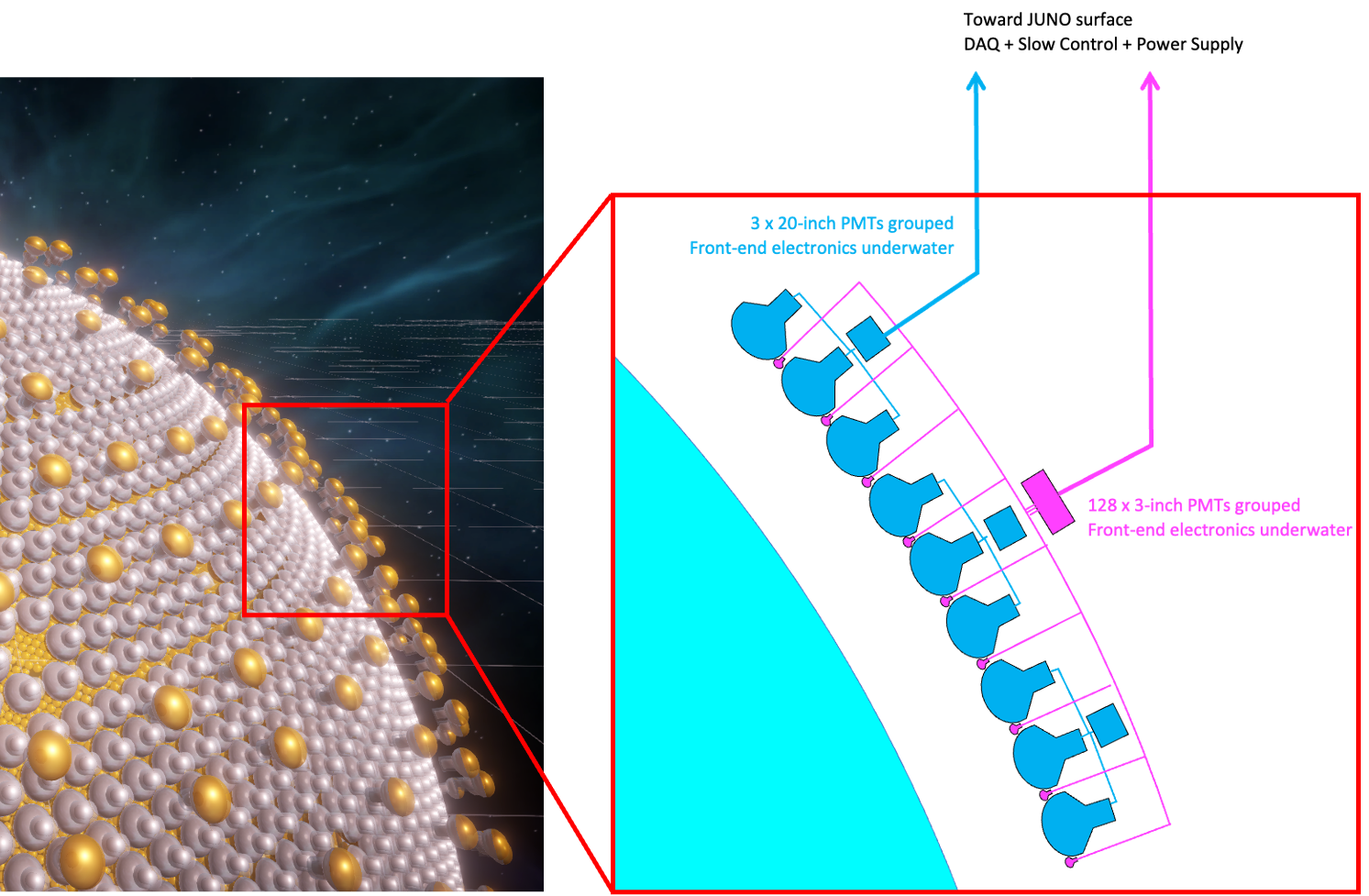}
    \caption[SPMT system overview]{\label{SPMT system overview} Cabling sketch of the photomultipliers with their front-end electronics underwater boxes. The 20-inch photomultipliers are grouped by 3, while the 3-inch photomultipliers are grouped by 128 over a maximum cable distance of 10 m. All electronics underwater boxes are then connected to the surface facilities (DAQ, Slow Control and power supply).}
\end{figure}

A maximum of 36,000 SPMTs was initially considered, based on the limited number of possible positions due to the stainless steel structure supporting the acrylic sphere of the central detector. However, additional constraints, including the cost of such a system and the limited benefit from additionnal SPMTs, led to a reduction in the number of SPMTs to 25,600. Reference~\cite{Li:2019hzc} provides in-depth information about the chosen 3-inch photomultipliers, their requirements, and their performance. The custom-made XP72B22, derived from the XP72B20 by Hainan Zhanchuang Photonics Technology Co., Ltd (HZC), was selected. This 3-inch photomultiplier, featuring ten acceleration stages, delivers excellent performance in terms of quantum efficiency (24.9\% on average), gain of 3$\times$10$^{6}$ at an average voltage of 1,100~V, single photoelectron (SPE) resolution (33.2\% on average), and dark count rate (500~Hz on average)~\cite{Cao:2021wrq}. Particular attention was given to the development of a low-radioactivity glass bulb.

Similar to the LPMT system, the SPMT readout electronics will be located underwater in watertight underwater boxes (UWBs). This design avoids transmission losses of analog signals over long cables to the surface and reduces the costs and risks associated with the required cabling. The constraints on PMT positioning and cabling drove the remaining requirements for the SPMT system. A schematic view of the positioning and cabling of the SPMT system is shown in Figure~\ref{SPMT system overview}.

\subsection{Granularity and cabling}

There are two primary considerations for the cabling design: balancing the maximum allowable cable lengths against the overall system reliability.
Signal cables of approximately 10 meters can effectively transmit low-amplitude, high-frequency signals with minimal degradation.
By defining coverage areas with a 10-meter radius for both cabling and signal readout electronics, each zone encompasses roughly 200 Small Photomultiplier Tubes (SPMTs).
Consequently, about 200 SPMTs would ideally be grouped and connected to a single underwater box for signal readout.

\begin{figure}[ht!]
    \centering
    \includegraphics[width=0.7\textwidth]{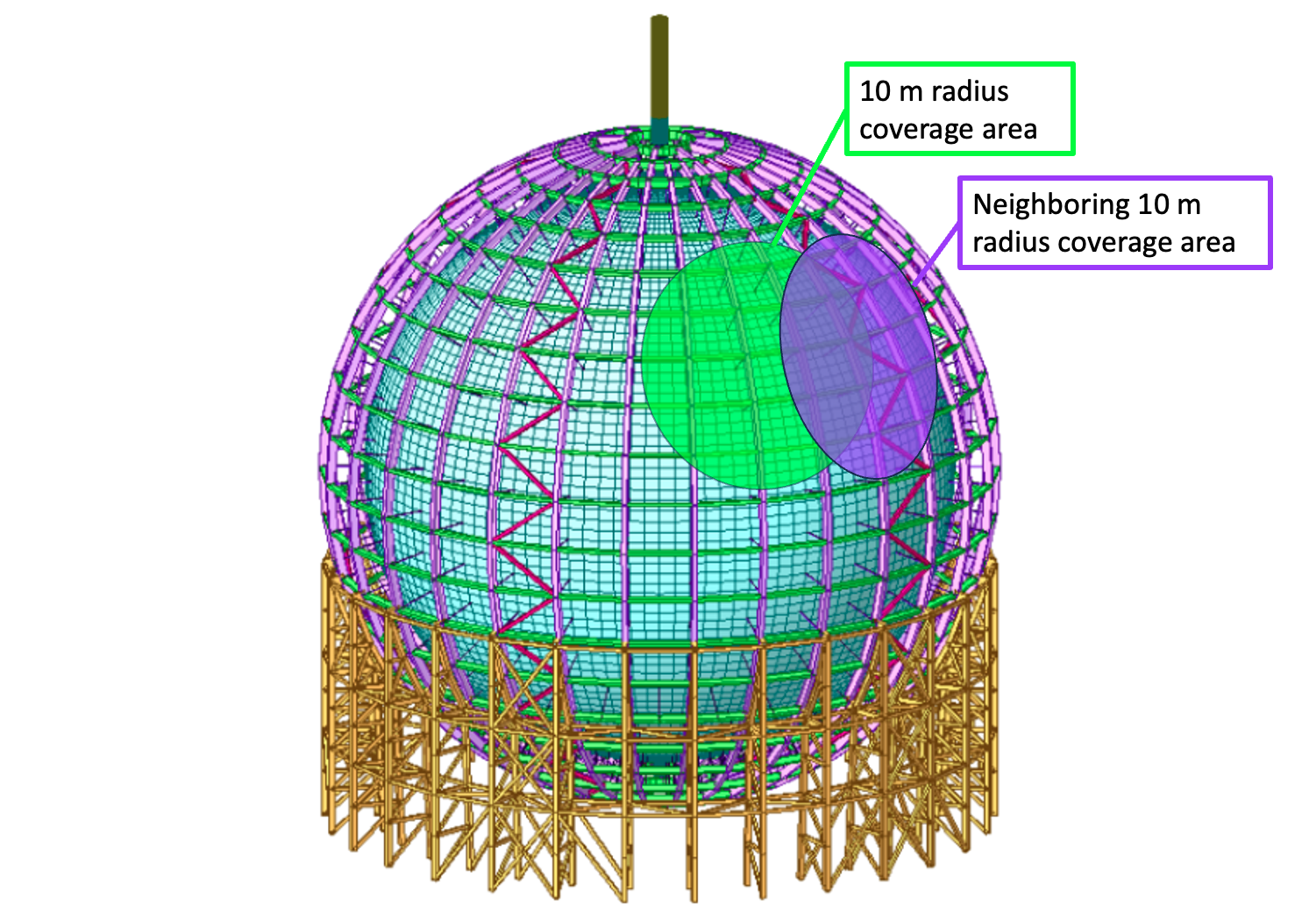}
    \caption[coverage]{\label{coverage} Illustration of the electronics coverage between neighboring central detector areas leading to a 50\% overlap redundancy for each area of the detector.}
\end{figure}

To enhance system reliability, a redundancy strategy is employed: a $\sim$50\% overlap in SPMT coverage is implemented across adjacent central detector areas, as illustrated in Figure~\ref{coverage}.
This redundancy ensures that if one readout electronics unit fails, the SPMTs connected to neighboring units prevent the creation of blind spots in the PMT coverage.
In a 10-meter radius zone containing $\sim$200 SPMTs, only 50\% of these SPMTs are allocated to a single readout unit.
This results in approximately 100 SPMTs per readout electronics unit.
However, 128 PMTs per underwater box is adopted as a practical compromise, given that channel readout typically scales as a power of 2.

Finally, each PMT requires both a high voltage (HV) power supply and a signal readout.
Using two separate cables for these operations would significantly increase the number of required cables, thereby raising the risk of failures at cable interfaces and escalating the overall cost of the SPMT system.
Given that transporting signals over a high-voltage power line and decoupling the signal from the HV at distances of 10--20 meters is a proven and reliable approach~\cite{Sato:2012jsa}, it was decided to power the PMTs and read out their signals using the same coaxial cable.
While this approach imposes stricter technical requirements on the cables, necessitating both high-voltage and radio-frequency specifications, it substantially reduces the number of required cables and interfaces.
A separate publication~\cite{Xu:2025czf} provides a detailed description of the photomultiplier instrumentation, including underwater cabling, connectors, and sealing strategies.
In this configuration, the high voltage must be generated underwater within the front-end electronics, using the low-voltage supply provided from the surface.
Consequently, the decoupling capacitor for the signal is relocated from the PMT divider to the front-end electronics.

\subsection{Front-End electronics specifications}

For the electronic readout of the PMT signals, the 16-channel CATIROC ASIC chip, designed at the OMEGA laboratory in France, was selected, as it satisfies the primary signal dynamics and timing constraints~\cite{JUNO:2020orn}.
Each CATIROC chip features 16 channels, enabling the readout 16 corresponding SPMT. Therefore, a single readout front-end board requires 8 CATIROC ASICs to handle the 128 associated SPMTs.
A total of 200 underwater boxes (UWB) and readout systems are thus needed to cover the 25,600 3-inch PMTs.

\begin{figure}[ht!]
    \centering
    \includegraphics[scale=0.5]{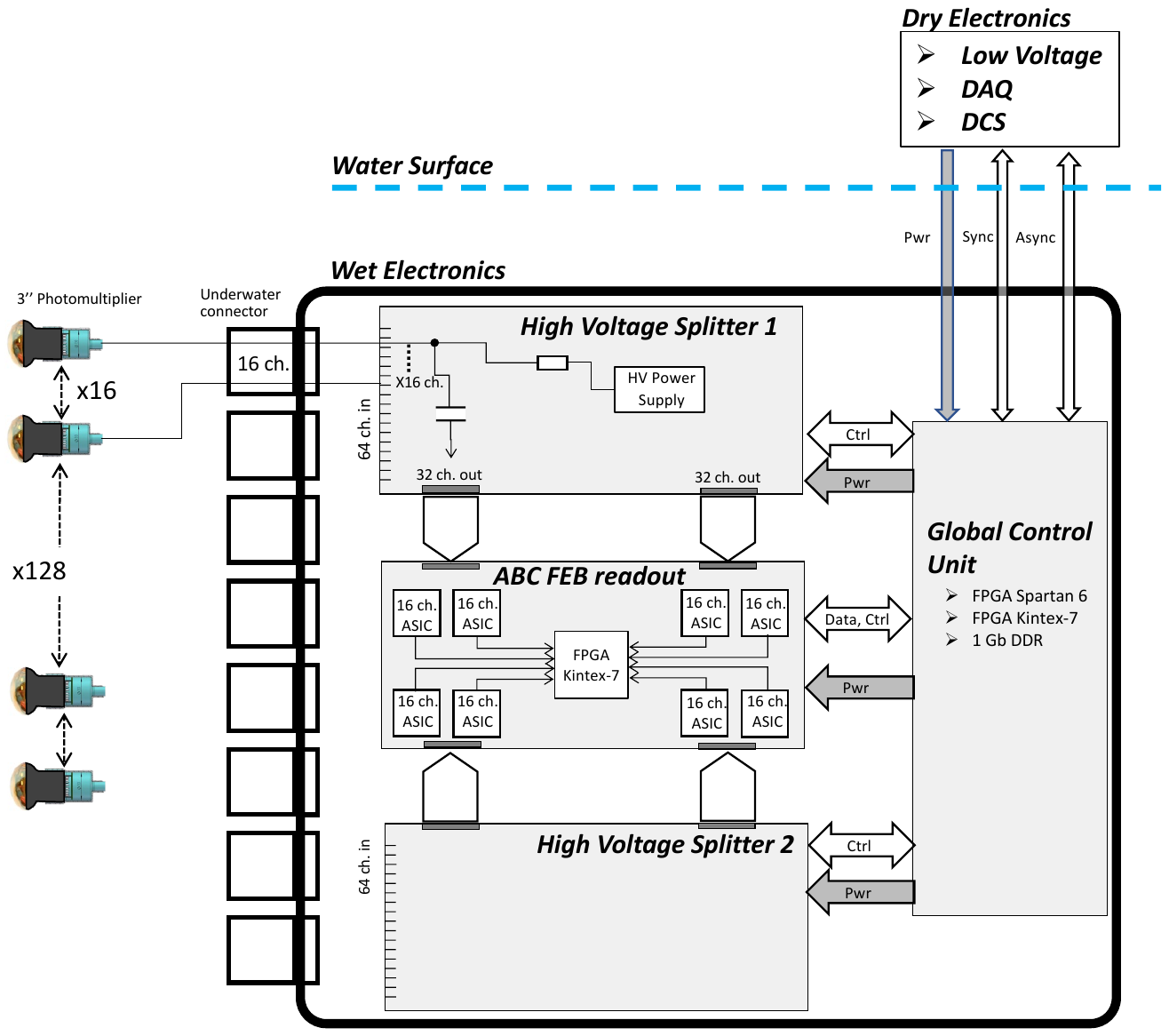}
    \caption[SPMT electronic system schematic for 1 underwater box]{\label{SPMT electronic system schematic} SPMT electronic system schematic for 1 underwater box. 2 High Voltage Splitter boards power the 128 PMTs and decouple their signal. The CATIROC ASICs on the ABC board process the PMT signals and the data is sent to the GCU board, which powers the whole electronic system, receives the clock signal from JUNO in a Synchronous mode (Sync) and transfers the data to the back-end electronics for Data Acquisition (DAQ) in an Asynchronous mode (Async). The Low Voltage (LV) supplies all the UWB.}
\end{figure}

The output signal from each SPMT is transmitted through the cable from its base to the readout electronics housed in the UWB.
The system comprises several electronic boards, each playing a distinct role in the data processing. A schematic overview of the SPMT electronics system is presented in Figure~\ref{SPMT electronic system schematic}.

The high-voltage power supply is integrated into the High Voltage Splitter board (HVS), which also decouples the signal from the high-voltage cable.
Each UWB contains two 64-channel HVS boards, responsible for supplying high voltage to all the connected 3-inch PMTs and decoupling the signals from the high voltage.
The decoupled signals are then forwarded to the front-end readout board, which hosts 8 CATIROC ASICs for signal processing and a Field-Programmable Gate Array (FPGA) for controlling each CATIROC and monitoring the board's environmental parameters.
The front-end readout board measures both the signal's arrival time and charge, transmitting this information via the FPGA to the Global Control Unit board (GCU).

\begin{figure}[ht!]
    \centering
    \includegraphics[width=0.9\textwidth]{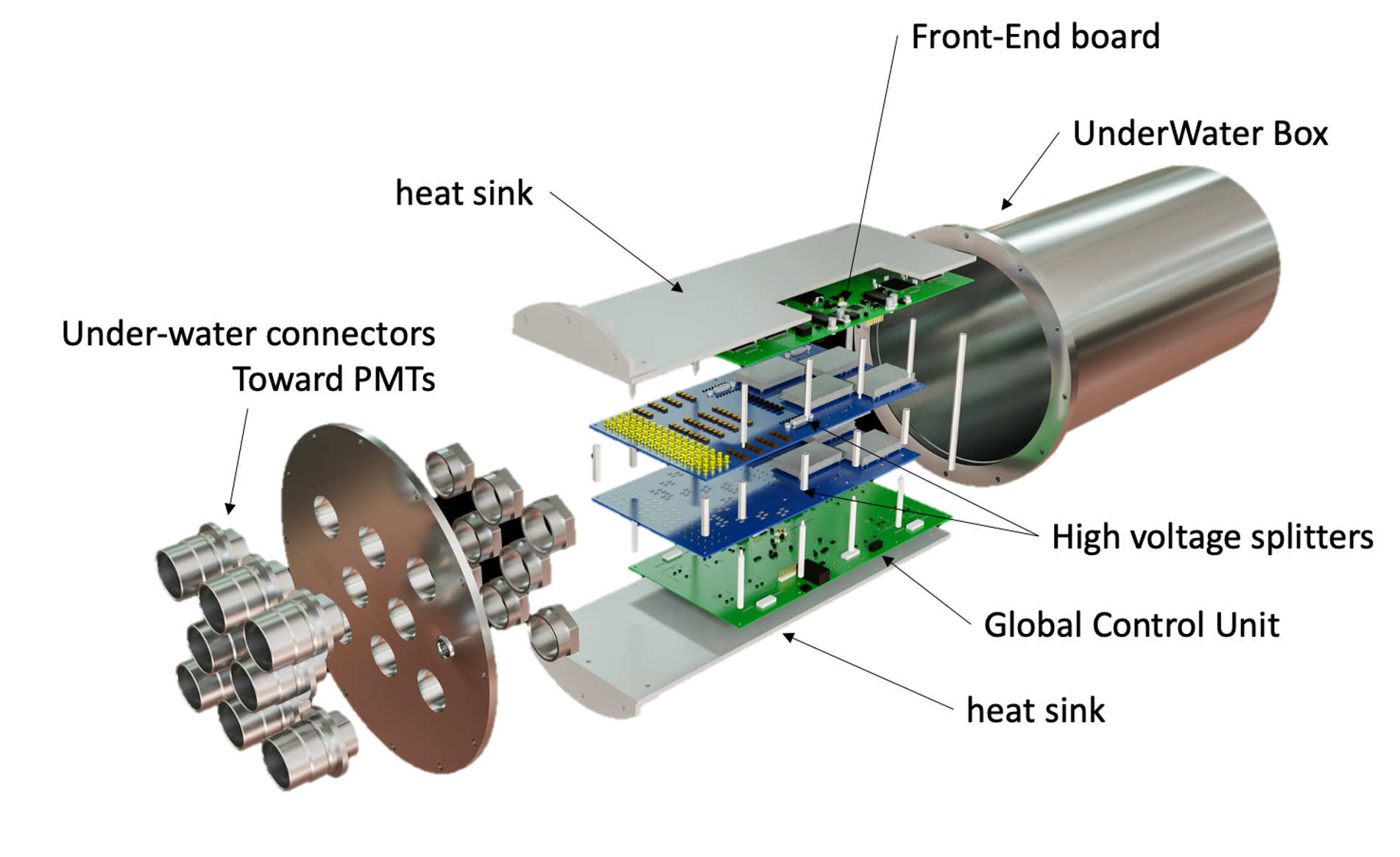}
    \caption[Exploted view of the SPMT electronic system inside underwater box]{\label{UWB scattered} Scattered view of SPMT electronic system inside underwater box. 16 PMTs per connector make up the 128 PMTs readout by the electronic system of one UWB. 200 UWB cover the 25,600  PMTs.}
\end{figure}

The GCU serves two primary functions:
it powers and controls the two HVS boards and the readout board, and it interfaces all controls and data with the surface electronics, including the Data Acquisition (DAQ), Slow Control (DCS) systems, and the low-voltage power supply system~\cite{SERAFINI2022167499}.
A comprehensive schematic of the system components connected to one of the 200 UWBs is shown in Figure~\ref{SPMT electronic system schematic} and Figure~\ref{UWB scattered}.
The following sections will detail each component of the SPMT system.

\section{The electronics architecture}

\subsection{The High Voltage Splitter board and High Voltage Unit}

\begin{figure}[ht!]
	\centering
	\includegraphics[width=1.0\textwidth]{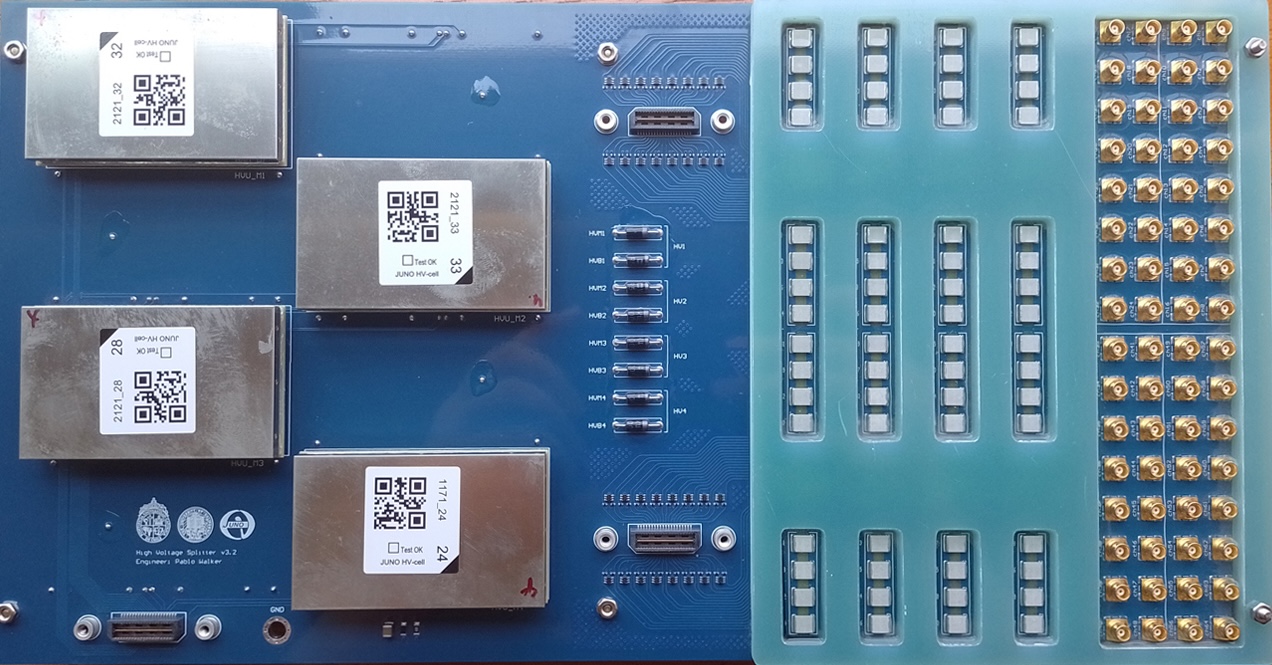}\vspace{1mm}
	\caption[The High Voltage Splitter PCB]{The High Voltage Splitter (HVS) board top side view. Glued to the right side of the board is an FR-4 piece with cutouts around the decoupling capacitors and the MCX connectors, used as a mold for pouring the insulating compound around these critical components.}
	\label{hvs_board}
\end{figure}

The High Voltage Splitter (HVS) \cite{WALKER2026171022} is the first board in the UWB read-out chain to interact with the PMTs. As previously explained, its functions include providing the high voltage necessary to bias the PMTs and decoupling the signal generated by the PMTs from the bias voltage. Each 128-channel UWB contains two HVS boards, with each board handling 64 channels. The board dimensions are 36.5~cm $\times$ 19~cm, and it is 4.1~mm thick. The Printed Circuit Board (PCB) has eight layers, with four layers dedicated to high voltage and signal routing, while the remaining layers are used for grounding. A picture of the board is shown in Figure \ref{hvs_board}.

\begin{figure}[ht!]
	\centering
	\includegraphics[width=0.6\textwidth]{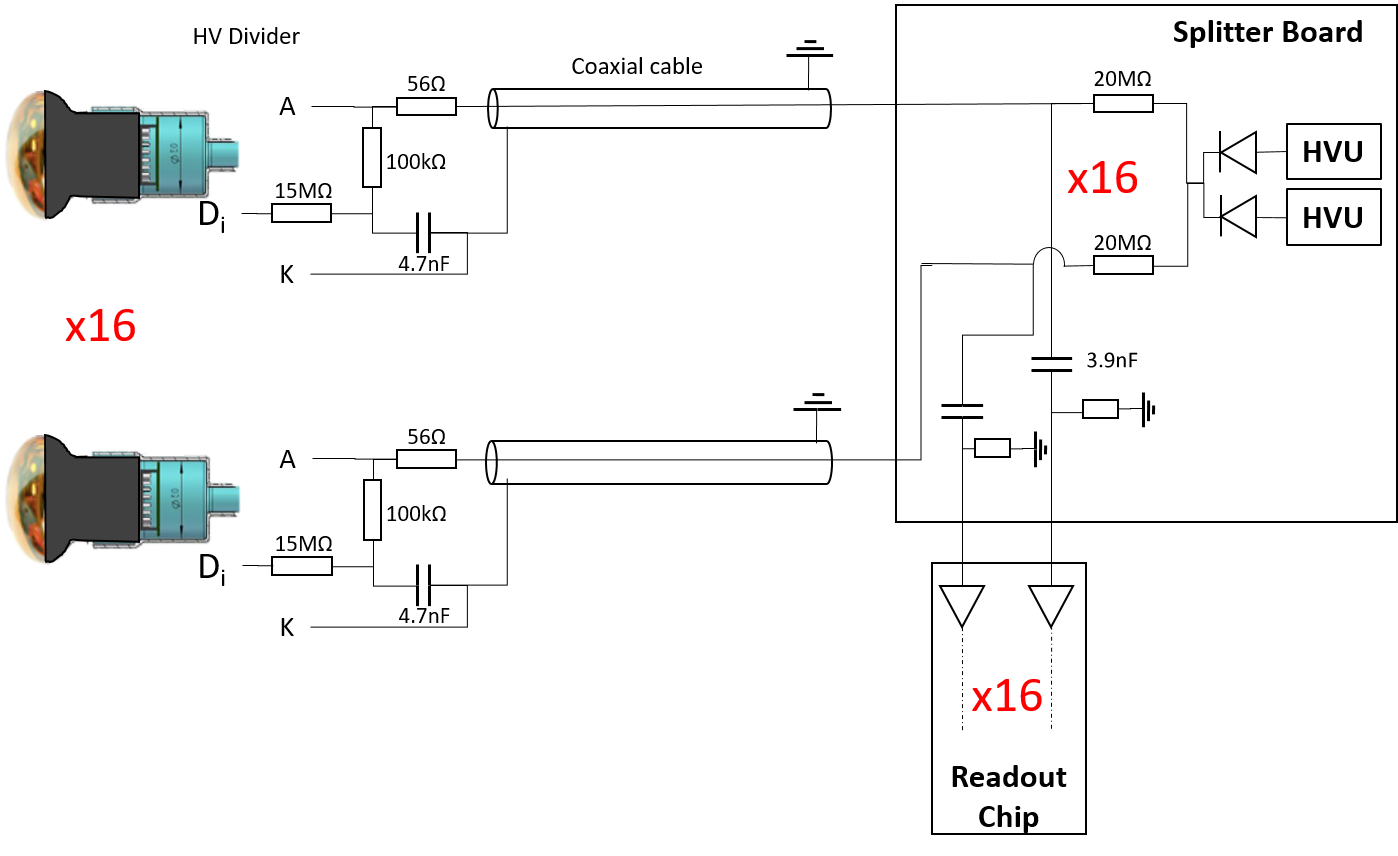}
	\caption[HVS simplified schematic]{Simplified circuit schematic for a 16-channel subgroup of the HVS board and the surrounding analog electronics. On each channel, a PMT connects to the HVS through a long coaxial cable, which carries both the high voltage necessary to power the PMT and the low voltage  signal.}
	\label{Decoupling schematics}
\end{figure}

The HVS houses eight copies of the High Voltage Unit (HVU), a custom made DC-DC converter circuit capable of generating the high voltage necessary to bias the SPMTs \cite{BELLATO2021164600}. The HVUs are powered by 24~V and have a configurable high voltage output which can be set anywhere within the range of 800-1,500 V.
A single HVU provides voltage to a set of 16 SPMTs simultaneously, with SPMTs in each set carefully selected to have very similar bias voltages at nominal gain. Under normal conditions, only four out of the eight HVUs on the HVS are used, while the remaining four serve as backups in case of failure, implementing a 1:1 redundancy scheme. The output of an HVU and their respective backup unit are connected together through diodes, capable of withstanding the HV reverse bias, as shown in Figure \ref{Decoupling schematics}. Although the HVUs are mounted on the HVS, they are powered and controlled by the GCU, through a single Q Series\textregistered High Speed 40-pin SAMTEC connector.

The high voltage output of an HVU is split into 16 channels using 20 M$\Omega$ series resistors, which also serve the secondary purpose of limiting the current draw from the HVU in the event of a channel failure, such as a short-circuit. The HVS uses custom hybrid SMD/THD MCX connectors from Axon\cite{axon:2024} to connect each SPMT, which are visible on the right-hand side of the board in Figure \ref{hvs_board}.

The signal is decoupled from the high voltage by 3.9~nF capacitors. The signals are then transmitted to the front-end board via two Q Series\textregistered High Speed 40-pin SAMTEC connector assemblies. The decoupling capacitors were carefully selected and extensively tested with special emphasis on the selection of the capacitance value to minimizing reflections and signal integrity, dielectric type to minimize capacitance variation due to applied voltage and aging, and long-term reliability metrics, such as failure in time (FIT).

\begin{figure}[ht!]
    \centering
    \includegraphics[width=0.65\textwidth]{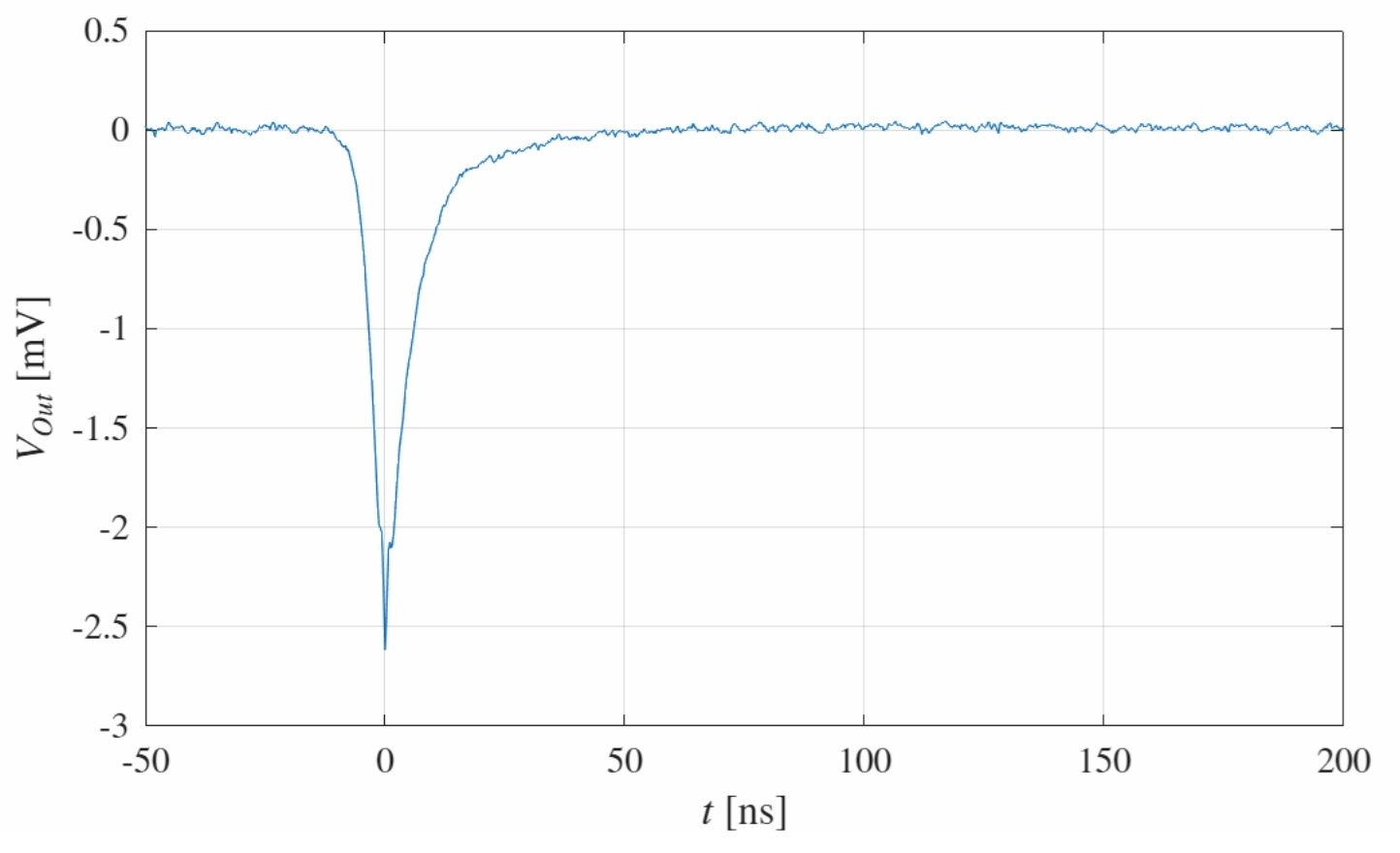}
    \caption[High Voltage Splitter SPE waveform]{Average SPE waveform measured on the output of the HVS while using a PMT connected with a 10~m cable at nominal gain ($3\times 10^6$). This plot represents the average of 1,000 captured SPE waveforms.}
    \label{High Voltage Splitter Spe waveforms}
\end{figure}

Figure \ref{High Voltage Splitter Spe waveforms} shows a representative plot of an average SPE signal at nominal gain, as measured on the output of the HVS, with the PMT connected through the 10m coaxial cable. The threshold of the trigger system on the ABC board is set to \(1/3\) of a photoelectron, which has an average amplitude of roughly 2~mV. Electronic noise and signal distortions, such as reflections and crosstalk, need to be limited to reduce the risk of false positives. On the analog part of the electronics it is thus required that the RMS noise and distortion levels must be kept below \(1/10\) of a photoelectron. An analysis of crosstalk between neighboring channels showed a bipolar pulse induced by PMT signals. This pulse peaked at a maximum of 1.6\% in the most affected channels. Although this is the limiting factor in the crosstalk budget of the front-end electronics, it remains well within system specifications.

Being the only board that handles high voltage inside the UWB, and given the stringent reliability constraints of the system, several steps were taken to ensure the long-term operation of the board. Critical components, including high voltage diodes, decoupling capacitors, and the fully populated PCB, underwent independent testing, with each item subjected to hundreds of hours of accelerated aging to characterize their reliability, with all of them complying with system-level specifications. At the layout level, carefully selected clearances were used to separate high and low voltage traces, and blind vias and thick dielectric layers were used to improve high voltage insulation while keeping the design relatively compact given the large channel multiplicity. Additionally, controlled-depth milling slots were included underneath the decoupling capacitors, and together with the base of the MCX connectors, were fully surrounded in a transparent, insulating compound called Pentelast-712 \cite{Penta:712}, as shown in Figure \ref{hvs_board}. Additionally, conformal coating was applied to all exposed high voltage pads.

\subsection{The ABC front-end readout board}

The front-end readout board, named ABC for ASIC Battery Card, plays a pivotal role in reading out, digitizing and packaging the signals from the 128 SPMTs, which arrive from the two HVS boards. The ABC board integrates eight 16-channel CATIROC ASICs and a high-performance FPGA (Kintex-7) on a 12-layers PCB, forming a critical component of the signal processing chain. The analog signals are received through four Q Series\textregistered High Speed 40-pin SAMTEC connector assemblies, with each connector managing 32 channels and interfacing two CATIROC ASICs. The board, measuring 35 x 17 cm, provides a compact and robust platform for signal processing.

Powered by a 12V supply, with specialized regulators splitting the main power into separate digital and analog streams. These power streams are distributed across different layers of the PCB, separating digital and analog components to minimize interference and optimize performance. Crafted from FR4 material, the 12-layer PCB of the ABC board ensures optimal signal integrity and interference mitigation. 
Critical components are strategically positioned on the board to optimize performance and ease integration. Warm components requiring specific cooling are located on the top layer, facilitating direct contact with a metallic heat sink. This arrangement ensures that temperatures are maintained below $40^\circ$C for the ASICs and $60^\circ$C for the FPGA during operation, enhancing their reliability and longevity. The board total consumption during operation is approximately 15 W,  with thermal dissipation ultimately transferred to the detector’s veto water. After component assembly, the PCB was coated with a conformal coating to enhance moisture resistance, prevent environmental degradation, and ensure long-term stability.

The FPGA provides a 160~MHz clock signal for the CATIROC ASICs, while digitized signals from the ASICs are clocked at 80~MHz, derived from the FPGA's clock signal.The two frequencies are locked together using a Phase-Locked Loop (PLL) implemented in the FPGA. To ensure synchronization, the lengths of all traces carrying digitized signals between the ASICs and FPGA are meticulously matched, as the CATIROC ASICs lack a dedicated clock output. The ABC board interfaces directly with the GCU for control, supply of power, initialization of the ASICs, and data output. The FPGAs (Kintex-7) on both boards are interconnected via a dedicated Q Series\textregistered High Speed 40-pin SAMTEC connector assembly. This comprehensive interface enables efficient control, data exchange, and firmware installation management. 

The ABC board offers two methods for the firmware installation and configuration. Direct programming via a JTAG connector on the PCB’s bottom side, used for factory firmware installation, and dynamic programming via the ABC/GCU interface for IPbus configuration during operation, facilitating real-time adjustments and maintenance.

To ensure reliability, the board utilizes a military-grade FPGA renowned for its robustness and durability. This strategic choice ensures maximum reliability, critical for applications demanding uninterrupted operation and longevity. The reliability of the ABC board has been validated through an accelerated ageing test. The board underwent continuous operation in a hot, controlled environment at $95^\circ$C for more than 500 hours. This testing demonstrated a Failure In Time (FIT) rate below 1900 (i.e., fewer than 1900 failures per billion operating hours), ensuring less than 10\% losses during the detector’s first six years of operation.

Standalone tests of the ABC board revealed the following performances: 
\begin{itemize}
    \item Mean RMS noise: ~2.3 Analog-to-Digital Converter units (ADCu) (10-bit ADC), with a signal-to-noise ratio of ~26 for a single photoelectron (PE). The electronic noise contribution (~4\%) is negligible compared to the 3-inch PMT charge resolution (~33\%).
    \item Charge linearity: Better than 0.05\% in the [0–10] PE range corresponding to the full dynamic range of the CATIROC ASIC in high gain.
    \item Time resolution: ~0.23 ns (well below the PMT’s Transit Time Spread of 1.6 ns).
    \item Crosstalk: Measured to be below 0.4\% across all 128 channels.
\end{itemize}

\begin{figure}[ht!]
    \centering
    \includegraphics[scale=0.14, angle=90]{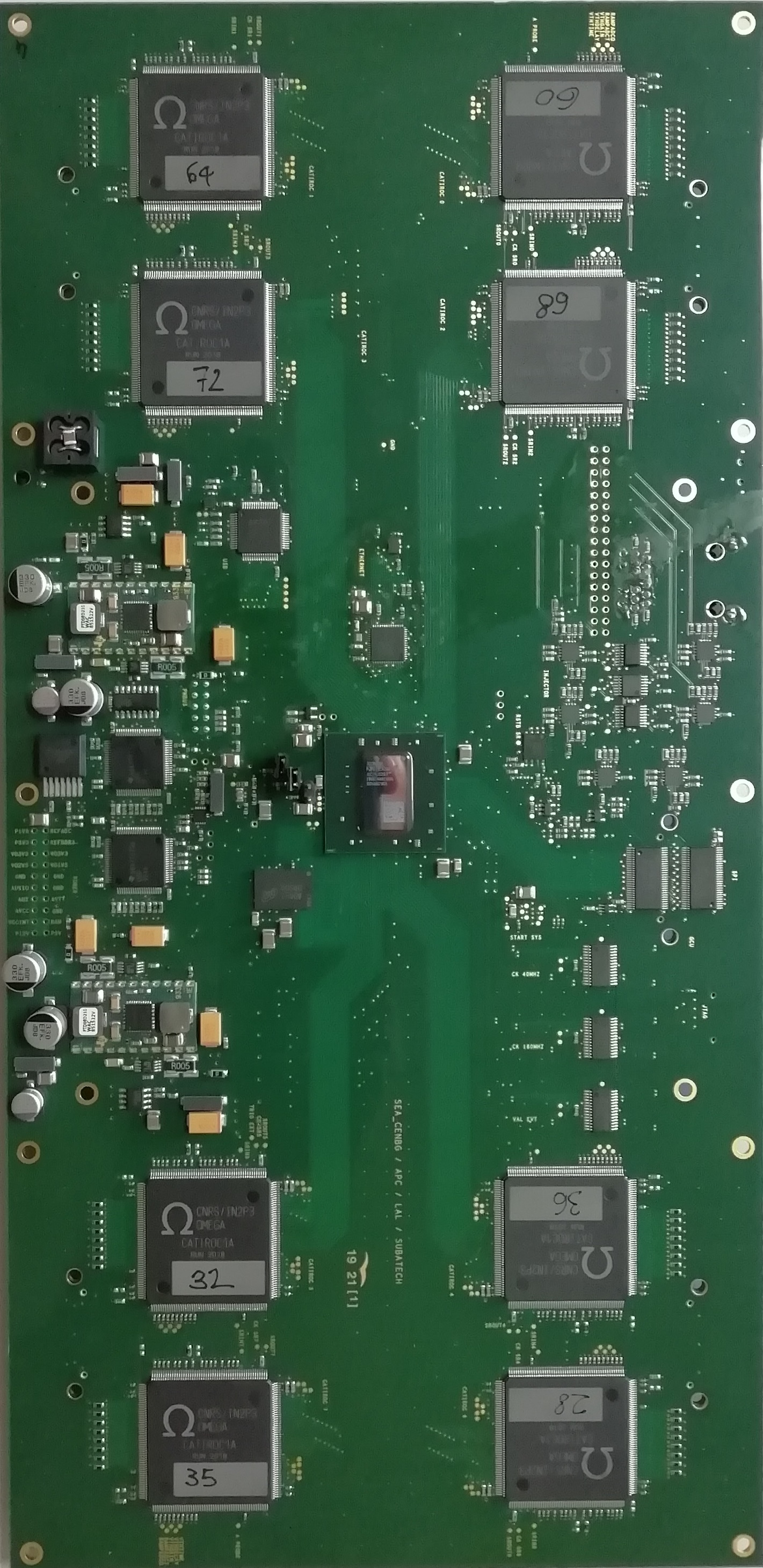}
    \caption[ABC V1.2 board]{ABC front-end readout board top side with main components. The 8 CATIROC ASICs are positioned on both sides of the board in a square formation and can be identified by the $\Omega$ symbol on their surface (OMEGA laboratory). The FPGA is the central piece of the board, as it collects all the data from the 8 ASICs. Other components are mainly decoupling capacitors. }
    \label{Views of ABC V1.2 board}
\end{figure}

\subsection{Global Control Unit}

After digitization, events are transmitted from the CATIROC ASICs to the FPGA, and then to the Global Control Unit (GCU) board via the Q Series\textregistered High Speed 40-pin SAMTEC connector. The GCU enables continuous data readout, packaging the data for transmission to the surface Data Acquisition (DAQ) system, where it is stored and analyzed.
In addition to managing data flow, the GCU plays a central role in initializing the SPMT system, supplying the synchronous clock received from the surface,  intialization the CATIROC chips and ensuring the firmware maintenance capability of the ABC. The GCU also powers and controls the High Voltage Units (HVUs) on the HVS boards, allowing for real-time monitoring and adjustment of the high voltage supplied to the PMTs. Finally, the GCU serves as the primary interface between the underwater electronics and the surface systems, facilitating communication with the DAQ, Slow Control (DCS), and power supply systems.

\begin{figure}[ht!]
    \centering
    \includegraphics[scale=0.8]{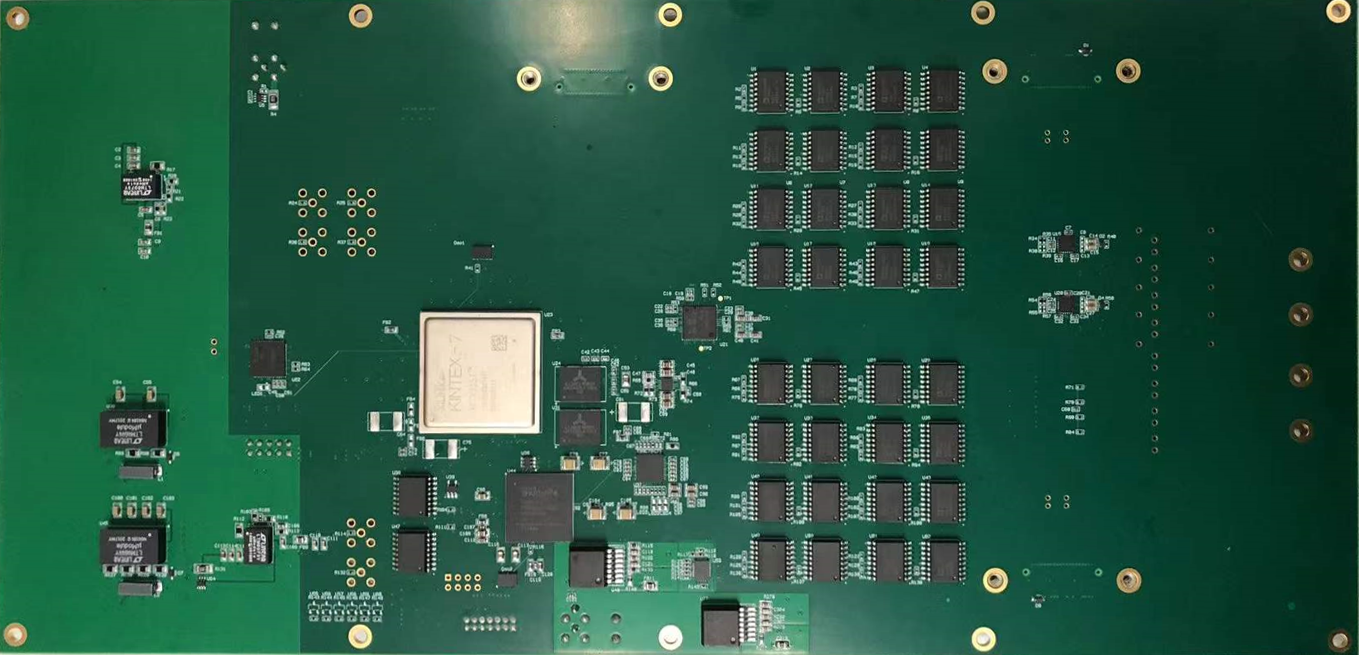}
    \caption[Global Control Unit]{\label{Global Control Unit} Global Control Unit board top side with main components, in particular the FPGA (Kintex-7).}
\end{figure}

It consists of a 35~cm $\times$ 17~cm PCB, with a thickness of 2~mm and comprises 16 layers. Eight of these layers are designated for signal transmission, five for grounding, and three to allow it to power the SPMT system. Its key components include a main FPGA (Kintex-7), a second FPGA (Spartan-6), and a 16 Gigabits DDR3. The GCU board is shown in Figure~\ref{Global Control Unit}.

It is powered at 36~V coming the surface, provides a 12~V power to the ABC board for communication, 24~V to the HVS boards, and the HVUs. Additionally, it supports safe remote reprogramming through a dedicated Spartan 6 FPGA for firmware updates.

As can be seen in Fig.~\ref{SPMT electronic system schematic}, a single GCU board locally connects to two HV splitter boards, one ABC board, and one pair of power cables. It ensures also the connection with the surface electronics, in particular to the Back-End Card (BEC) via a synchronous link through FTP cables for clock distribution and trigger management, to a Gigabit Ethernet switch, connected itself to the DAQ system, via an asynchronous link through UTP cables for DAQ and slow control, and to an AC-DC power supply via low resistance power cables.

\subsection{The Underwater Electronics Box}

\begin{figure}[ht]
\begin{minipage}[b]{.48\textwidth}
\centering
\includegraphics[width=1\textwidth]{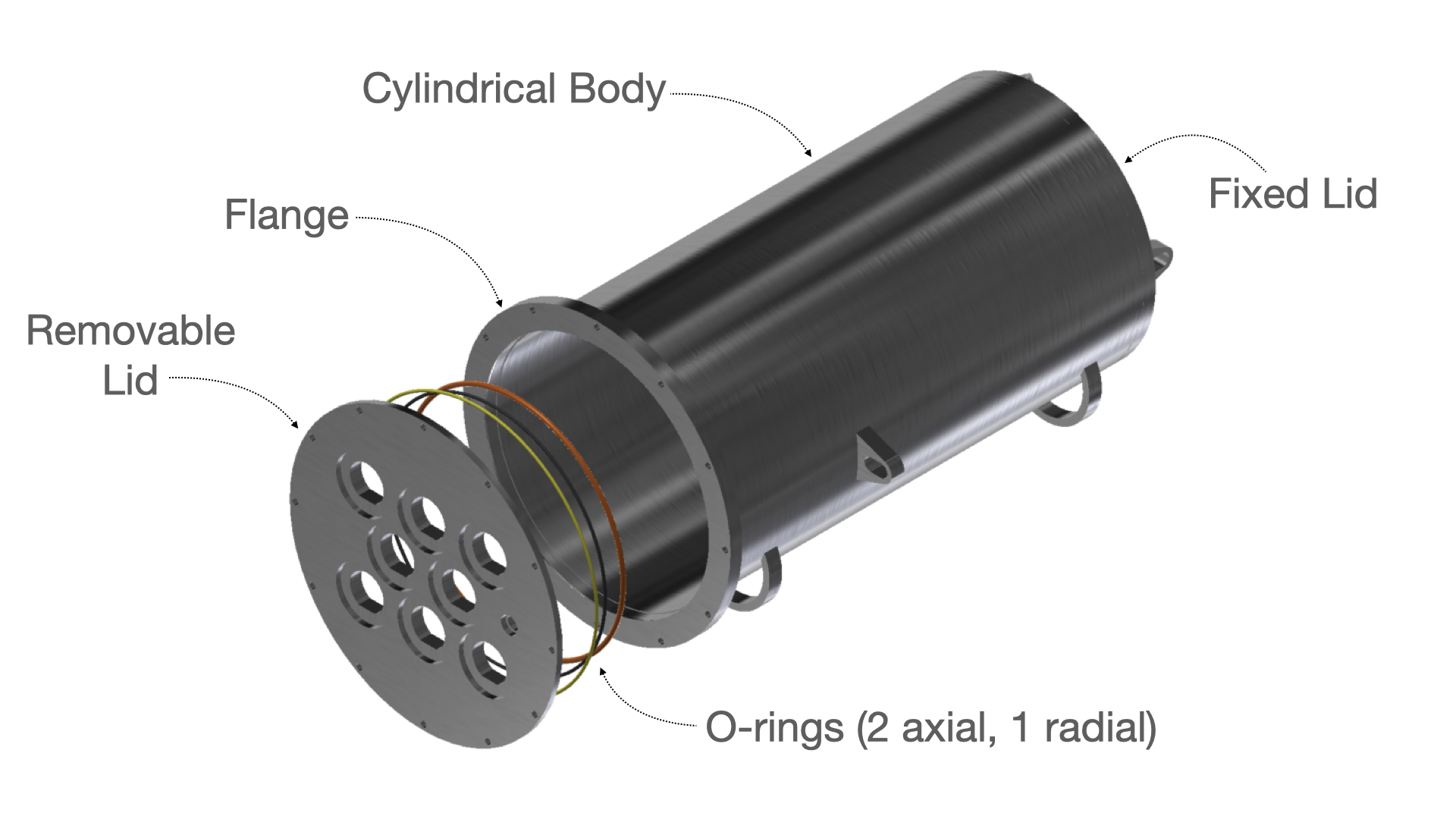}
\caption{Exploded view of the UWB and its components}
\label{UWB}
\end{minipage}
\hfill
\begin{minipage}[b]{.48\textwidth}
\centering
\includegraphics[width=1\textwidth]{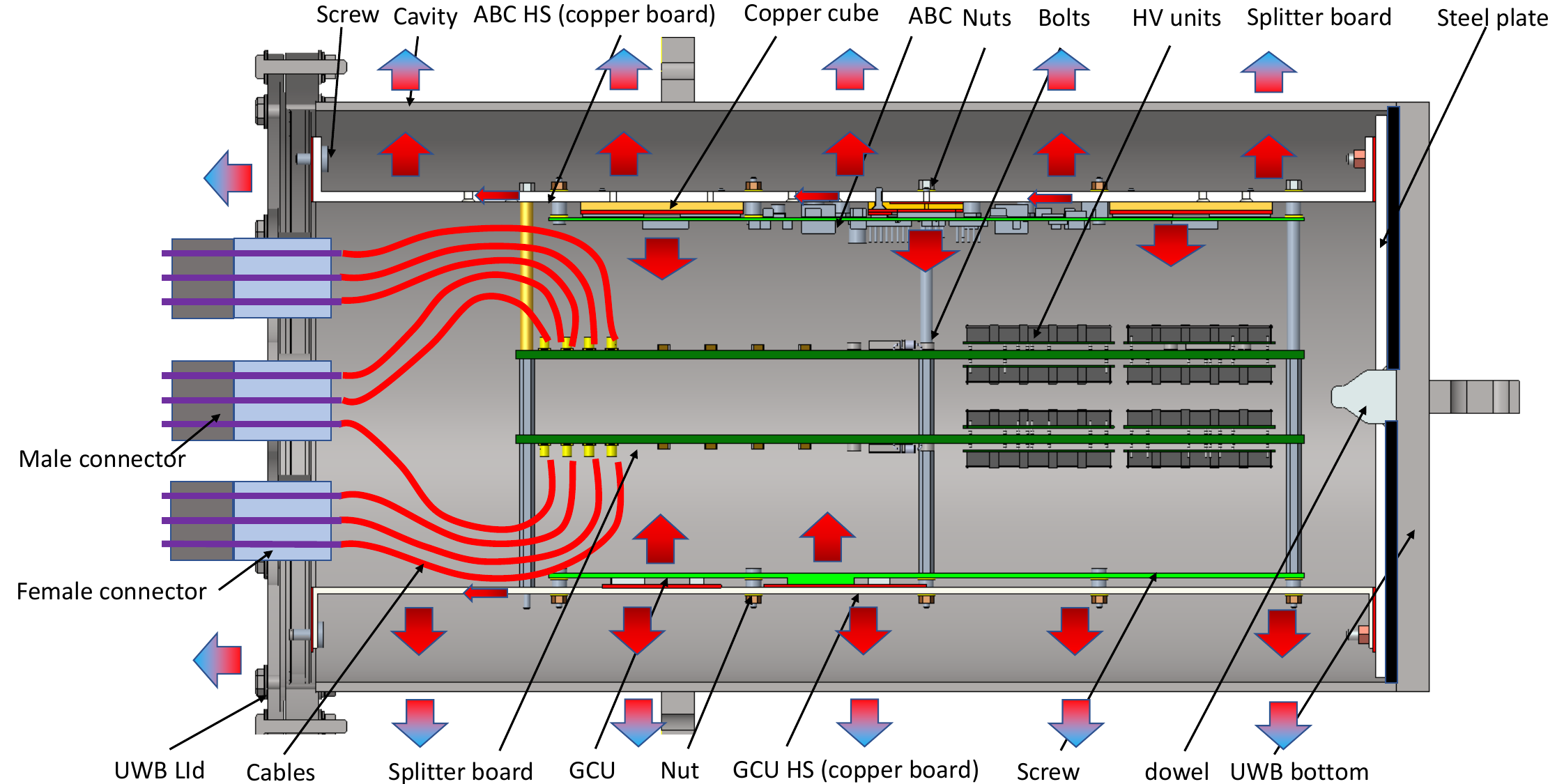}
\caption{Sketch of the thermal management of the front-end electronics in the Underwater Box based on thermal conduction and convection}
\label{HeatSink}
\end{minipage}
\end{figure}

\indent \textbf{Design: }Each set of electronics, able to operate 128 3” photomultipliers, is mounted in an independent underwater electronics box, the UWB. Thus, 200 underwater boxes are required for the SPMT system.\\
Each UWB consists of a cylindrical body, a flange, and two lids, one removable and one fixed. The body is a 50~cm long (internal measurement), 273~mm wide with a maximum eccentricity of 1mm over the internal diameter, 4.15~mm thick, stainless steel (SS304L) pipe. The flange is a 15~mm long cylinder with roughly the same internal diameter as the body but carefully machined to achieve a circular cross-section with nearly no eccentricity. The fixed lid is a 15~mm thick disk that is welded to the opposite end of the body. The removable lid, which is 20 mm thick, is secured with screws on the periphery and sealed using three O-rings made of different materials. Two axial O-rings and one radial O-ring provide redundant sealing to ensure water tightness. The UWB was simulated with the Finite Element Method (FEM), using a convergence criterion of 5\% or less.
The removable lid holds 8 custom Axon underwater connectors\cite{axon:2024}, each handling 16 channels, through which the PMTs are connected. The three main cables connecting each UWB to the surface electronics (Low voltage power, synchronous and asynchronous connections) fit in a long metallic bellow also welded to the lid. 
Four handles are placed in the body for manipulation and installation, as well as one hoisting ring in the fixed lid. 
An exploded view of the UWB and its components is shown in Figure~\ref{UWB} and also in Figure~\ref{UWB scattered}.

\indent \textbf{Manufacturing Procedure: } At first, the fixed lid was welded to the cylinder, followed by the flange. All welding was done using Tungsten Inert Gas (TIG), initially with internal fusion welding, followed by a second external layer using filler material to eliminate the possibility of leaks. The flange was then finely re-machined using Computer Numerical Control (CNC) technology to compensate for any deformations that occurred during welding to achieve flat surface of sealing. The removable lid was very carefully machined using a combination of CNC milling and Electrical Discharge Machining to ensure proper sealing with the O-rings and the underwater connectors. The bellows were welded on the removable lid after verifying with 3D imaging that no deformations occurred in the process. Acceptance was achieved through a combination of methods: visual inspection, measurement of key dimensions using calipers, roughness meters, and customized instruments for quick verification of compliance with critical parameters, as well as sealing tests conducted with a helium leak detector. 

\indent \textbf{Reliability: }The reliability of the UWB without connectors or bellows was tested with a custom-made high-pressure tank able to withstand pressures up to 30 bar in water warmed at 90°C. 
Nine mechanical models of UWBs each with two removable lids were manufactured using the same materials and protocols but with a shorter cylinder length plus a full-size UWB were placed simultaneously in the tank to increase statistics. As a first long-term test, the nine UWBs were subjected for 60~days to a constant pressure and temperature of 8~bar and 60°C, respectively. No leaks were detected and the water did not go past the first o-ring in all cases, achieving a FIT number of 726, lower than the requirement of 1,900. A second, more extreme long-term test was performed where the 9 UWBs were operated with only one o-ring per removable lid in the three possible positions for a total of 1,800 hours (2.5 months). The pressure was constantly set to 8~bar and the temperature to 77°C. No leaks were observed in any case. This was equivalent to subjecting the o-rings to about 35 years of aging, and resulted in a FIT of 127.   

\indent \textbf{Thermal management: }The power consumption of the SPMT front-end readout electronics system is primarily driven by the high-power chips on the ABC and GCU boards, with the ABC board consuming approximately 15~W and the GCU board consuming about 5~W. To ensure long-term stable operation of the electronics system, a cooling structure is essential for maintaining the temperature of each chip below 60$^\circ$C in a water environment of 21~$\pm$~0.5$^\circ$C.
To achieve this, a thermal management system based on thermal conduction and convection has been designed, as illustrated in Figure~\ref{HeatSink}. The system features a multilayer sandwich structure comprising two copper heat sinks, the GCU, the ABC, and the splitter boards. This assembly is secured to the UWB lid using four screws. The high-power chips, each consuming more than 0.5~W, are soldered onto the surface of the GCU and ABC boards, directly facing their corresponding heat sinks.
To optimize heat transfer, screws, nuts, gaskets, and insulating washers maintain a precise gap of 1.4~mm to 1.7~mm between the power chips and the heat sinks. This gap is filled with thermal silicone rubber, a soft, elastic heat transfer medium with a thermal conductivity of 4~W/(m$\cdot$K). The rubber is compressed to 70--85\% of its initial thickness, ensuring excellent thermal contact between the chips and the heat sinks. During operation, heat generated by the high-power chips is efficiently transferred to the heat sinks, where it is dissipated.

\begin{figure}[ht]
\begin{minipage}[b]{.48\textwidth}
\centering
\includegraphics[width=1\textwidth]{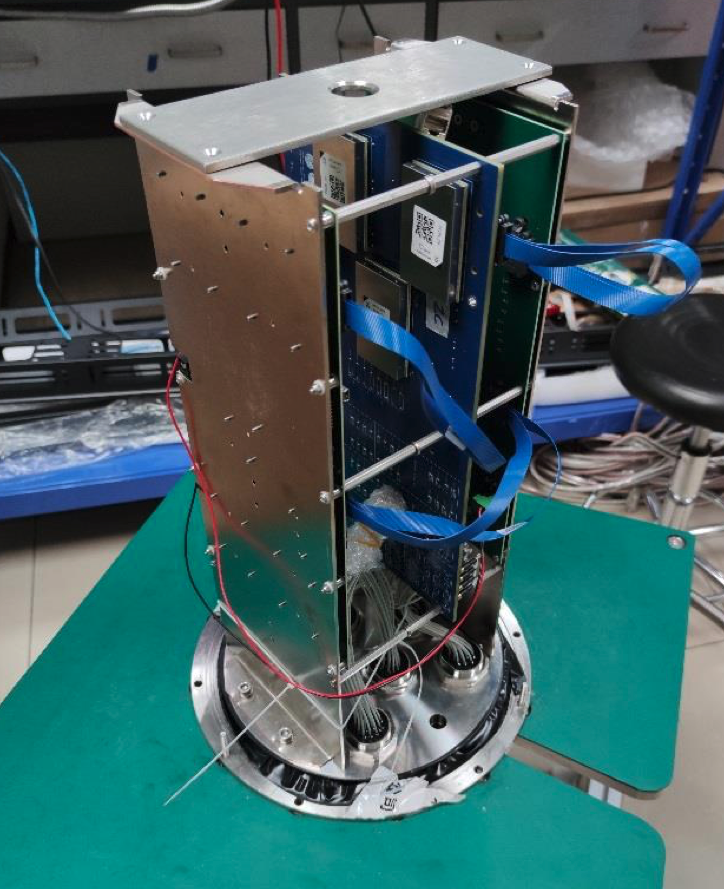}
\caption{Picture of the electronics stack with the heat sinks, mounted over the lid of the under water box.}
\label{fig:uwbstack}
\end{minipage}
\hfill
\begin{minipage}[b]{.48\textwidth}
\centering
\includegraphics[width=1\textwidth]{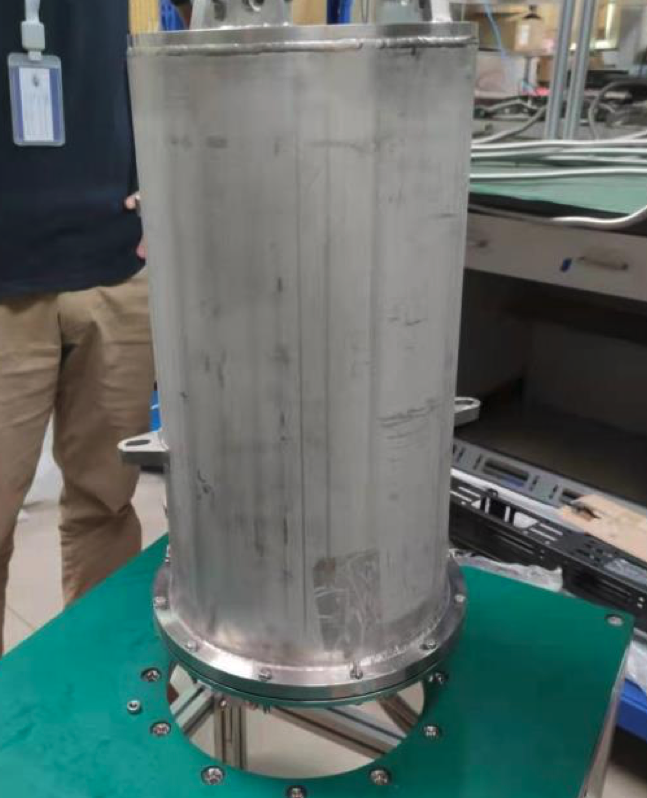}
\caption{Under water box after closure.\newline \newline}
\label{fig:uwbclosed}
\end{minipage}
\end{figure}

Several studs maintain a 60~mm separation between the ABC and GCU boards, ensuring easy integration and natural air convection to prevent heat buildup. A stainless-steel plate with a central hole secures two heat sinks near the UWB's bottom, with a dowel at the center providing radial support. This design prevents cantilever effects, protecting the electronics boards from stress and vibration during transport or when positioned at an angle. Figure~\ref{fig:uwbstack} and Figure~\ref{fig:uwbclosed} show a complete stack of electronics boards assembled together with their heat sinks over an UWB lid, and the same stack enclosed in the body of the UWB.

\subsection{Slow Control and Global Firmware architecture}

\begin{figure}[ht!]
    \centering
    \includegraphics[width=0.98\textwidth]{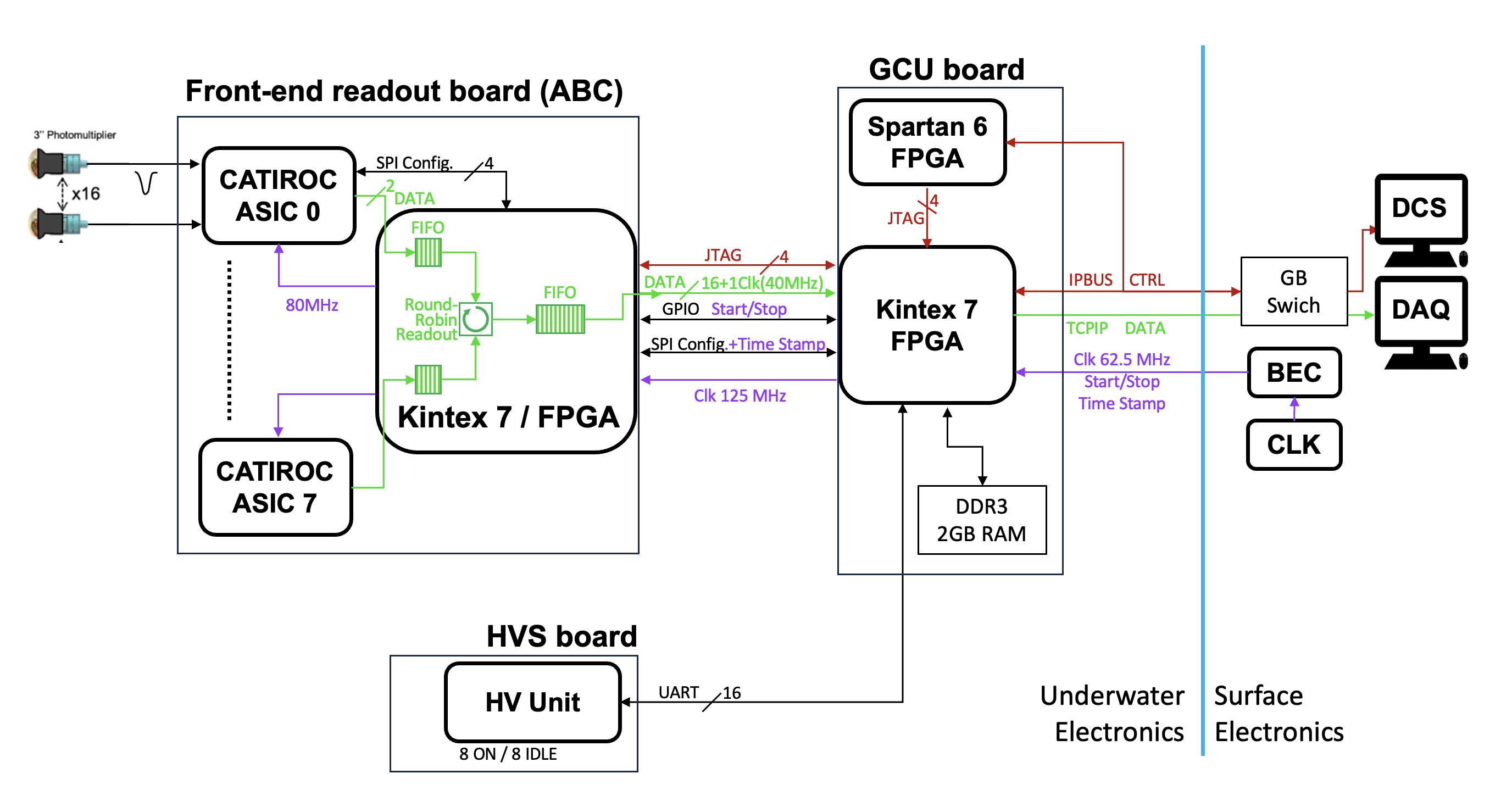}
    \caption[HWlinks]{\label{HW_links}Scheme of the electronics links between the key components of the SPMT system.}
\end{figure}

The SPMT electronics system uses a firmware architecture that manages the flow of data from signal capture to transmission, ensuring the control and synchronization across the 200 UWB. The firmware is distributed across the ABC (front-end readout) and the GCU (control unit), each with specific responsibilities within the system as illustrated in Figure~\ref{HW_links}. Together, these functionalities support continuous data acquisition, configuration, and timing necessary for the PMT readout. The primary functionalities of the firmware are organized as follows.\\
\indent \textbf{System Start-Up and Firmware Uploading:} At start-up, the system relies on a factory firmware embedded in the Spartan 6 FPGA on the GCU board. This embedded firmware establishes initial communication with the surface electronics and enables access to the FPGA (Kintex-7) on the GCU board. The FPGA (Spartan 6) configures the FPGA (Kintex-7) as a master JTAG controller. The dedicated protocol between the GCU and the ABC includes JTAG lines allowing the configuration of the ABC's FPGA (Kintex-7) from GCU's FPGA (Kintex-7) configured as JTAG master. IP address assignment is done using a RARP-type process managed by the IPBus protocol. Each Spartan 6 will have its own address, enabling large-scale deployment and allowing the 200 underwater electronics modules to be programmed in parallel.\\
\indent \textbf{CATIROC ASIC Configuration:} The CATIROC ASIC requires the configuration of 328 Slow Control (SC) parameters. This sequence of data is first sent to the GCU board’s FPGA (Kintex-7) via an IPBUS protocol. The GCU's FPGA (Kintex-7), acting as an SPI (Serial Peripheral Interface) Master, communicates with the ABC’s FPGA (Kintex-7) and writes the configuration to each of the eight CATIROC ASICs.\\
\indent \textbf{Clock Distribution:} A system clock of 62.5 MHz is provided to all GCUs through the Back-End Card (BEC) on the surface. This clock is received by the FPGA (Kintex-7) on the GCU, where it is used to generate a 125 MHz master clock along with other necessary frequencies for firmware operation. This 125 MHz clock is then sent to the ABC, where it serves as the primary timing reference for the ABC firmware.\\
\indent \textbf{Time Synchronization:} With clock distribution in place, precise timestamp synchronization is ensured across all detectors via the White Rabbit protocol. Upon power-up, a calibration and synchronization phase between the BEC and GCU cards ensures that all GCUs have the same initial timestamp, provided by the BEC. At each acquisition start, the synchronized timestamp is transmitted from the GCU to the ABC via a high-speed SPI protocol (125 MHz). The timestamp is embedded in the data format on the ABC and then retransmitted to the GCU, providing consistent timing data across the entire system.\\
\indent \textbf{High Voltage Control on the HVS Boards:} The system incorporates control of the high voltage supplied by the High Voltage Units (HVU) on the HVS boards. The GCU board manages this aspect through a dedicated UART protocol (Universal Asynchronous Receiver/Transmitter), allowing both status monitoring and settings adjustments of the HVU.\\
\indent \textbf{Data Aquisition and Transfer:} During operation, each CATIROC ASIC sends data to the FPGA (Kintex-7) on the ABC board, which reads these inputs in a round-robin mode by a FSM (Finite State Machine) implemented in its firmware. The ABC's FPGA assembles the data into a FIFO buffer, which is then transferred to the GCU's FPGA (Kintex-7) FIFO. From there, data is transmitted to the Data Acquisition (DAQ) system via a TCP/IP protocol. The data transmission between the ABC and GCU boards occurs through a dedicated 16-bit protocol, sending 16 bits every 3 clock ticks (i.e., 37.5 ns). 

\subsection{Radiopurity}

The radiopurity control of all the materials is of great importance to achieve a very low radiogenic background for the physics goals in JUNO. This background dominates the single hit rate in the detector. The PMTs and their readout electronics contributes critically due to the radioactivity of the glass (for PMTs) and the PCB. Although they are separated from the LS by a $\sim$2~m water buffer attenuating the gamma flux. Detailed Monte-Carlo simulations have demonstrated that the selected materials will lead to a single rate lower than 10 Hz considering both an energy threshold $E_{th}$ of 0.7 MeV and a fiducial volume (FV) cut of 0.5 m in the LS \cite{JUNO:2021kxb}. For the SPMT system, the requirement on the single rate from the 25,600  3-inch PMTs and their readout electronics is 0.2~Hz.

\begin{table}[ht!]\small
\centering
 \caption{ \label{tab:radiopurity} Summary of the radiopurity screening of the main SPMT electronic readout components. The activity of each component in $^{40}$K, $^{238}$U chain and and $^{232}$Th chain has been divided by 128 channels and is expressed in mBq/channel. The last column corresponds to the expected single rate of each component in the FV with $E_{th}\geq$0.7 MeV for all the 200 UWBs using Monte-Carlo simulations and experimental activities.}
 \smallskip
	\begin{tabular}{c|c|c|c|c|c}
    \hline
    \textbf{UWB component} & \textbf{Mass} & \textbf{$^{40}$K} & \textbf{$^{238}$U chain}  & \textbf{$^{232}$Th chain}  & \textbf{Single rate in FV}  \\ 
             & [kg] & [mBq/ch] & [mBq/ch] & [mBq/ch] & [mHz] \\ 
    \hline 
    UWB structure & 31.30 & 67 & 20 & 14 & 1.4 \\ 
    \hline 
    HV Splitter boards  & 1.93 & 278 & 228 & 281 & 25.7 \\ 
    \hline 
    ABC front-end board  & 0.35 & 15 & 19 & 26 & 2.3 \\ 
    \hline 
    GCU board & 0.35 & 128 & 17 & 20 & 1.5 \\ 
    \hline 
    Heat sink & 2.5 & <5 & <4 & <12 & <0.4 \\ 
    \hline 
    \textbf{Total} & 35.0 & 508 & 304 & 361 & \textbf{31.3} \\ 
    \hline     
    \end{tabular}
\end{table}

To fulfill this requirement for the electronic components, the main materials of the SPMT front-end electronics in the UWB listed in Table~\ref{tab:radiopurity} have been carefully screened by low background gamma spectroscopy. The activity of $^{40}$K as well as $^{238}$U and $^{232}$Th chains have been converted in mBq/ch for each component of the UWB. The last column gives the contribution per component to the single rate in the fiducial volume with $E_{th}\geq$0.7 MeV for all the 200 UWBs. Finally, the total single rate due to the radioactivity of the SPMT readout electronics is about 0.031 Hz. Combining with the contribution from the PMTs (0.050 Hz), the HV divider (0.046 Hz) and the potting (0.022 Hz), the total single rate of the SPMT system is about 0.15 Hz in compliance with the 0.20 Hz requirement, the contribution from the readout electronics being only 21\% from the total.

\section{Performance}

The SPMT electronics system is designed to efficiently read out pulses generated by the 3-inch PMTs, achieving a threshold sensitive enough to trigger on single photoelectrons.
Its primary functions include measuring the charge and arrival time of these pulses, digitizing the information, and transmitting it with high bandwidth.
Key performance criteria---such as low noise, high linearity, and negligible dead time---are prioritized to meet the stringent requirements of the experiment.
To validate these performance metrics, multiple systems comprising 128 photomultipliers were connected to the SPMT electronics using identical configurations in terms of cabling, connectors, and mechanical integration.

\subsection{Noise and Cross-talk}

\begin{figure}[ht!]
    \centering
    \includegraphics[width=0.8\textwidth]{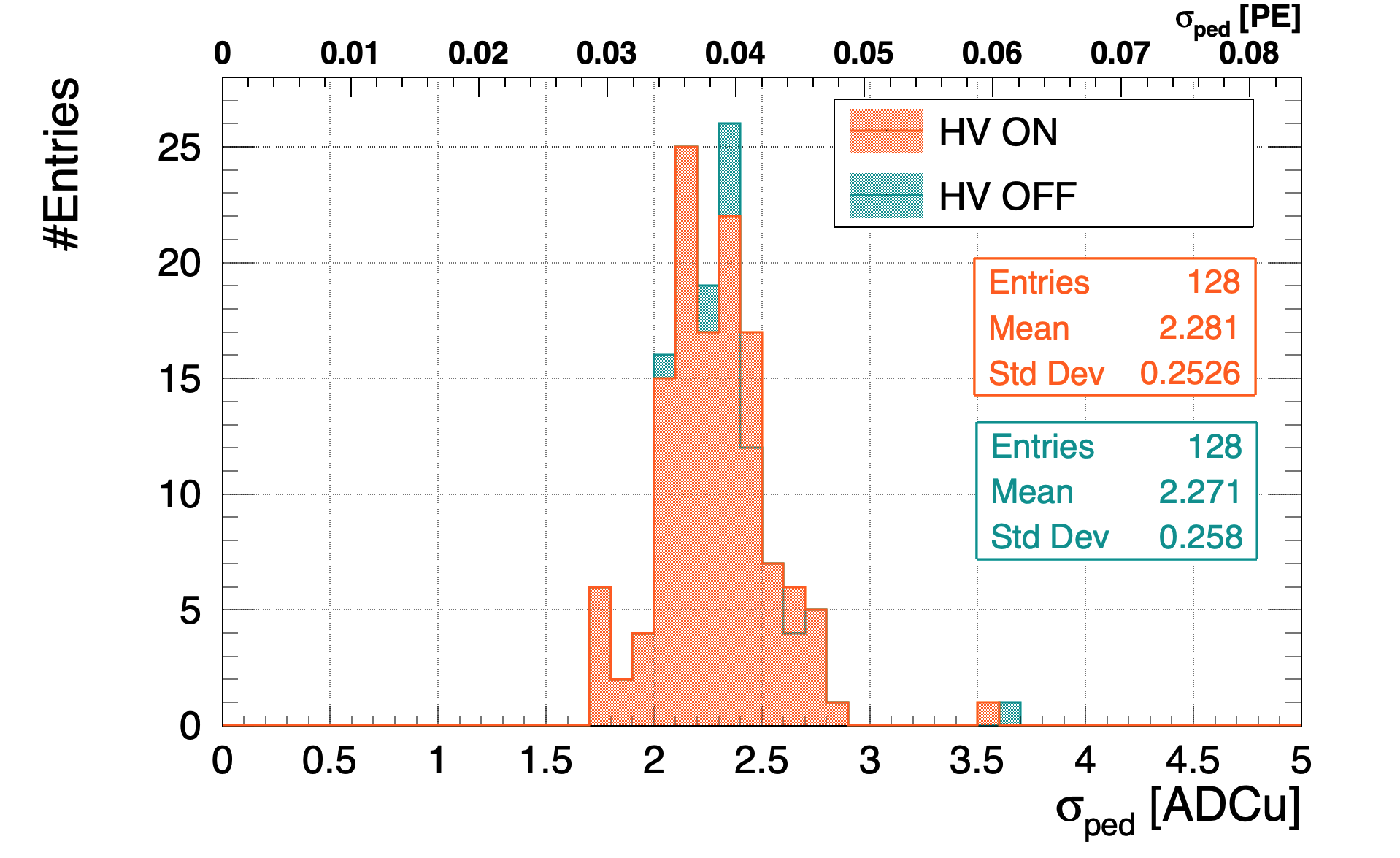}
    \caption[ChvsQ]{Pedestal RMS distribution of the 128 channels without (HV OFF, green) and with HV applied to PMTs (HV ON, orange). The level of noise $\sim$2.3 ADCu is consistent with the noise of the ABC front-end board. In addition, no additionnal noise is observed when PMTs are powered on}
        \label{Sigma Pedestal}
\end{figure}

The noise characterization involves a dedicated trigger, periodically generated by the ABC's FPGA (Kintex-7). This trigger prompts the CATIROC ASICs to perform readouts, enabling the measurement of pedestal charge values. The distributions have been fitted with a Gaussian and the resulting standard deviations ($\sigma_{Ped}$) across the 128 channels are illustrated in Figure~\ref{Sigma Pedestal}.

  The mean noise (or standard deviation) when PMTs are not powered (green) is measured 2.27 ADCu which is fully  consistent with the average RMS value observed in standalone operation of the ABC. A second pedestal standard deviation distribution, recorded with nominal High Voltage values applied to PMTs (orange), showed a mean noise value of 2.28 ADCu fully consistent with the previous measurement. This confirms that the overall system, comprising all boards powered in their operational configuration, does not add substantial noise compared to the standalone ABC. With the standard CATIROC preamplifier gain of 20 intended for operation during the experiment, the observed mean noise level \(\sigma_{elec}\) of 2.3 ADCu is equivalent to 0.04 PE, i.e. 4\% of the charge of a single PE. 
  When combining quadratically with the 3-inch PMT resolution \(\sigma_{PMT}\) of 33.2\%, \(\sigma_{elec}\) contributes less than 0.7\% to the SPE PMT resolution. These results confirm the trigger threshold, set at a third of a photoelectron, and suggest the capability of lowering the threshold to enhance trigger efficiency.

 To evaluate crosstalk between channels in the full SPMT system, a single 3-inch PMT (source channel) was powered with high voltage (HV), while the remaining 127 channels were physically disconnected. A single LED, flashing at approximately 1~kHz, was positioned above the source channel to serve as the sole light source. The LED-induced signals in the source channel were easily identifiable due to their high charge.
 The analysis focused on detecting hits in the disconnected neighboring channels that occurred within 40~ns of a hit in the source channel. Crosstalk was predominantly observed in two adjacent channels and showed a strong correlation with the charge of the source channel. Specifically, crosstalk became detectable when the source channel’s charge exceeded approximately 8 photoelectrons (PE) and reached a near-100\% probability when the charge was around 42~PE or higher.
The charge ratio between the neighboring channels and the source channel was consistently measured to be at or below 0.4\% in the two channels exhibiting the highest crosstalk. For all other monitored channels, this ratio was either zero or significantly smaller, demonstrating effective isolation between channels.

\subsection{Digitization and PMT performances}

\begin{figure}[ht!]
    \centering
    \includegraphics[width=0.8\textwidth]
    {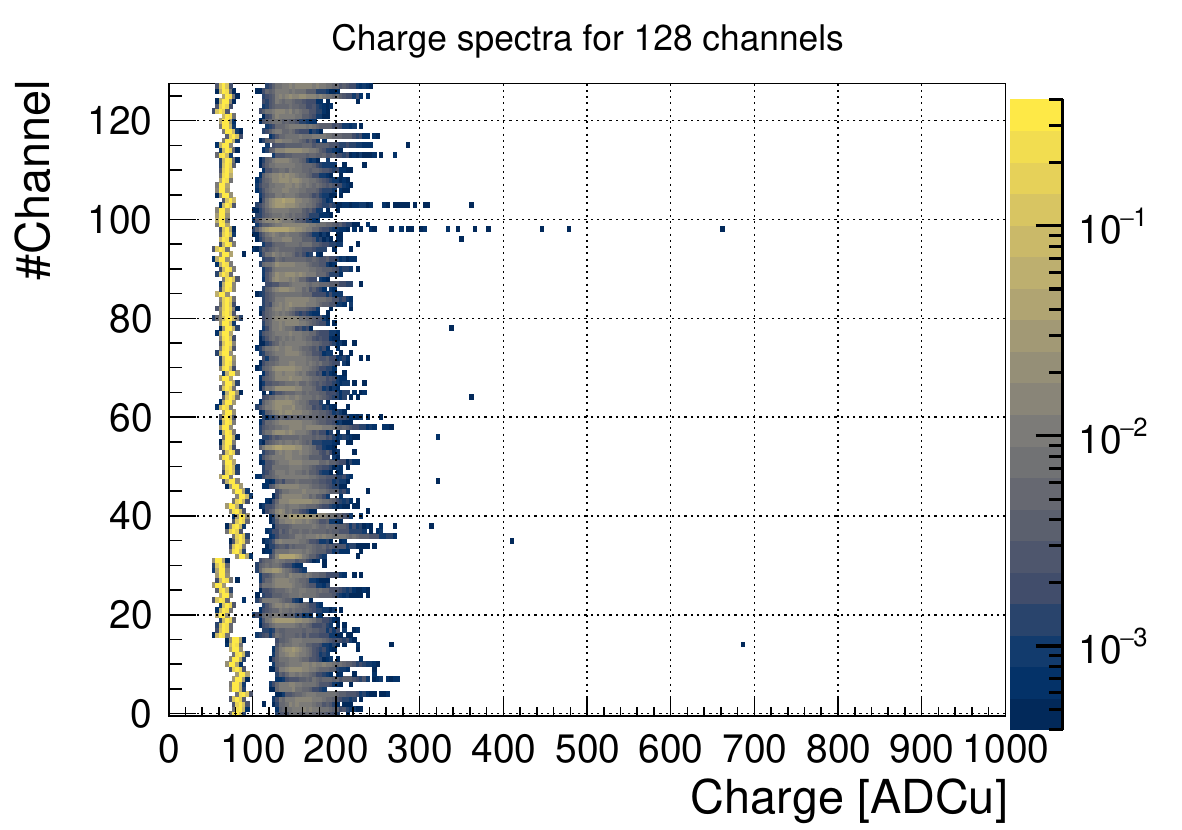}
    \caption[128 SPE Spectrum]{Charge spectra of the 128 PMT channels. The arbitrary spectrum intensity is normalized for each channel to account for difference of illumination between PMTs. Both single photo-electron (between 100 and 200 ADCu) and pedestal (lower than 100 ADCu) peaks are visible and well-separated.}
        \label{128 SPE Spectrum}
\end{figure}

\begin{figure}[H]
    \centering
    \includegraphics[width=0.8\textwidth]
    {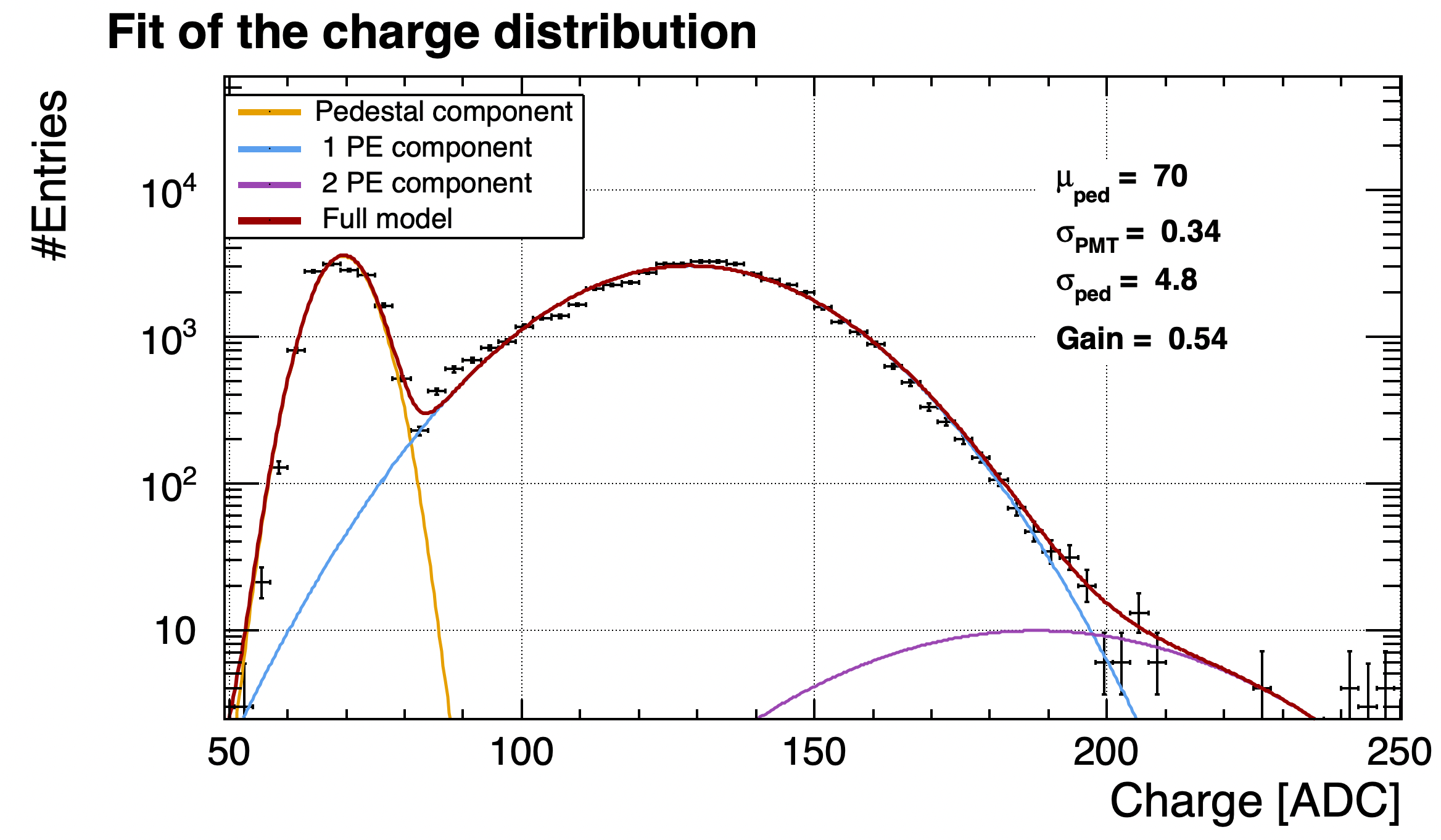}
    \caption[SPE Spectrum]{Example of the charge from a 3-inch PMT after the full SPMT readout electronics. The Single PhotoElectron spectrum and the noise are clearly visible. The fit extracts the PMT features, highlighting a charge resolution of 34\% for a gain of 0.54 pC/PE (3.38 $\cdot 10^{6}$), which aligns with the average nominal resolution of approximately 33\% across the 25,600  PMTs in the system.}
    \label{SPE Spectrum}
\end{figure}

The charge spectra of 128 SPMTs placed in a dark room are presented in Figure~\ref{128 SPE Spectrum}. One can clearly observe the SPE peak well separated from the pedestal. An illustration of the charge spectrum for a given channel is shown in Figure~\ref{SPE Spectrum}. The gain of the PMT is at 3$\cdot$10$^6$ for which the PMT is expected to have a resolution of $\sim$33\%. The fit performed on the whole spectrum allow to extract a PMT resolution of 34\% consistent with expected value. The impact of the overall front-end electronics noise is thus negligible.   
The overall gain of the detection chain, from the PMT to the ASIC amplifier, is approximately 0.5 pC/PE. The CATIROC ASIC operates with a high-gain dynamic range of 0–5 pC (0–10 PE), and switches to a low-gain range of 5–60 pC, extending the measurement up to 120 PE.

\subsection{Deadtime and bandwidth}

As established in Ref.\cite{JUNO:2020orn} , the CATIROC ASIC introduces two intrinsic sources of dead time: (i) a trigger dead time of 60–90 ns following a first PMT signal on a given channel, and (ii) a digitization dead time when more than two PMT signals occur within a 6–9 \textmu s window on the same channel. These limitations are intrinsic to the ASIC and cannot be mitigated at the firmware level. Additional losses may arise if the downstream data transfer bandwidth is saturated.

Beyond these intrinsic limitations, the bandwidth of the full system depends on both the hardware architecture and the firmware. The data chain involves two main boards: the ABC and the GCU. Digitized data are captured and processed in the onboard FPGA (Kintex-7), which builds events and stores them using a structured multi-FIFO architecture. 
Data are transmitted from the ABC to the GCU through a dedicated hardware link composed of 16 parallel buses operating at 26.7 MHz with a 16‑bit word width. With the current data format, this corresponds to 80 bytes transferred every 175 ns, yielding a maximum sustained throughput of 57 MB/s. This value has been experimentally validated using logic probes and oscilloscope measurements.

A toy-model was developed to estimate the system’s bandwidth limits based on realistic conditions. It incorporates the internal FIFO sizes and organization in both boards, the number and type of interconnections, and the latencies involved at each step of the data flow. This calculation shows that the firmware architecture and the multi-FIFO data management are not the limiting factors. Instead, the primary bottleneck comes from the hardware link between the ABC and the GCU at 57 MB/s.

This conclusion was confirmed experimentally by configuring the ABC FPGA with an internal trigger capable of firing all 128 channels simultaneously at rates up to 625 kHz. Saturation of the data flow was observed at approximately 36 kHz per channel, corresponding to an aggregate bandwidth of 57 MB/s, in agreement with the limitation of the ABC–GCU hardware link. No additional dead time or data loss is observed beyond this limit, demonstrating the robustness of the firmware and validating the bandwidth model.

\subsection{Supernova Resilience Assessment}

\begin{figure}[H]
    \centering
    \includegraphics[width=0.9\textwidth]
   {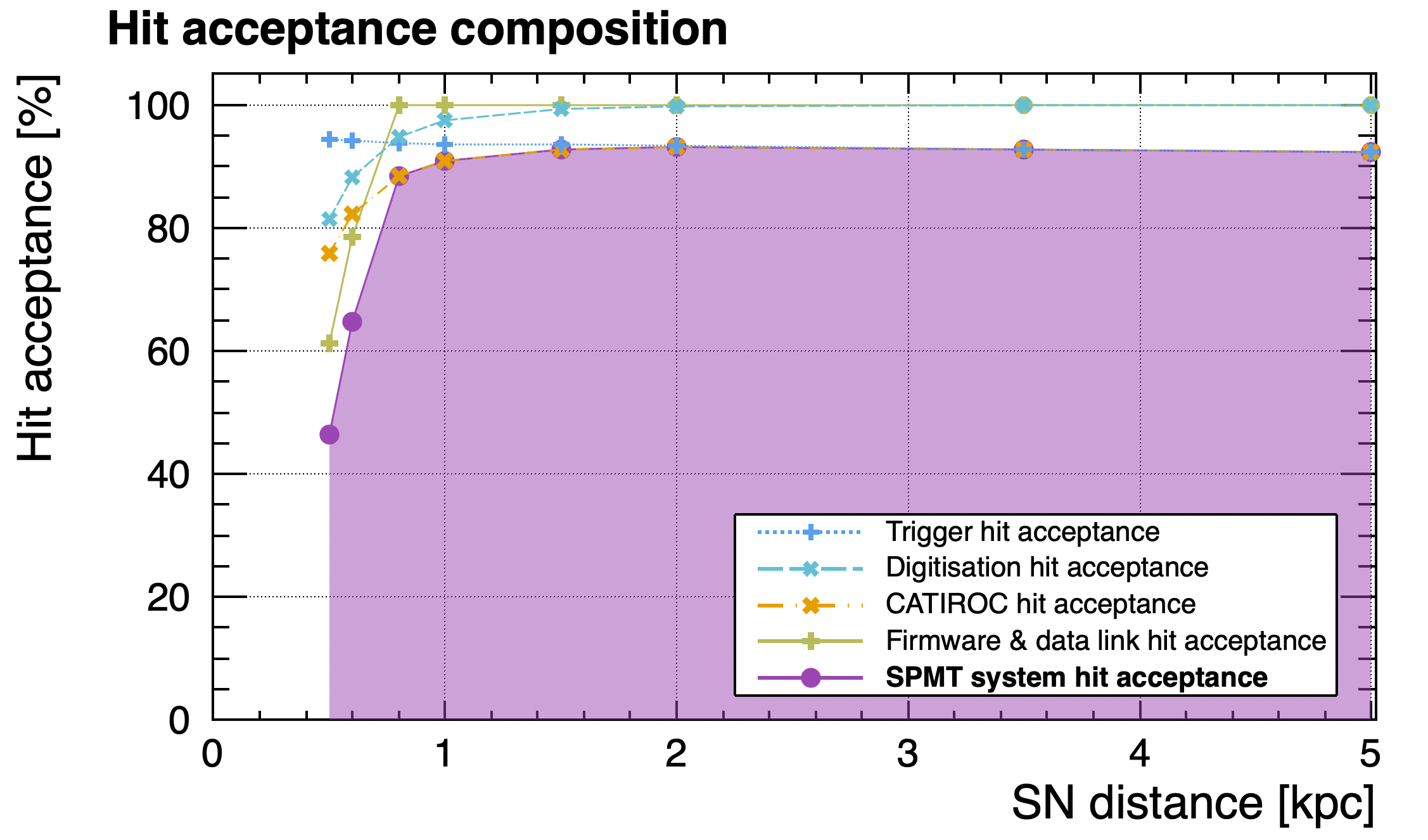}
    \caption[SN Data loss]{Small PMT system hit acceptance for a 20-solar-mass CCSN. A hit is defined as a scintillation photon detected by a PMT, hence a neutrino interaction results in numerous hits. A detailed description of the legend is provided in the main text.}
    \label{SN Data loss}
\end{figure}

The bandwidth toy-model is further used to evaluate the behavior of the readout system during high-rate transient events such as a Core-Collapse Supernova (CCSN). The Figure ~\ref{SN Data loss} illustrates the SPMT's hit acceptance for a 20-solar-mass CCSN, taking into account all relevant hardware constraints, inluding the acceptance of CATIROC's trigger (dotted blue line) related to the trigger dead time and the acceptance of CATIROC's digitization process (dashed light blue line) related to the digitization dead time. These two acceptance form together CATIROC's hit acceptance (dash-dotted orange line). In addition to CATIROC's acceptance, the hardware limitation causes a bandwidth acceptance (solid green line). All of the above form together the SPMT system acceptance (purple line).

For distances below approximately 0.8 kpc, the dominant limitation arises from the ABC–GCU bandwidth. At larger distances, the system bandwidth is no longer limiting and the acceptance is instead governed by the intrinsic CATIROC dead times. In this regime, the full electronics chain—from digitization to GCU buffering—preserves essentially 100\% of the accepted hits. Even at a distance of 1.0 kpc, the overall hit acceptance remains above 90\%, demonstrating the system’s ability to handle the burst rates expected from a nearby supernova.

It is worth noting that this simulation currently models the electronics pipeline up to the GCU board, without accounting for limitations on the downstream Data Acquisition (DAQ) system. To extend the evaluation beyond the GCU's data transfer, a dedicated assessment was performed to estimate the occupancy of the GCU’s onboard DDR3 memory in the event of a Supernova burst. Based on the expected hit rate for a CCSN at 1 kpc, approximately 1.5 million hits per ABC are anticipated during a 0.7 s burst, accounting for ASIC dead times and ABC-GCU transfer limitations. Depending on the data packing efficiency, this corresponds to a data volume between 19 MB (best case, full 8-hit packing) and 44 MB (worst case, one hit per packet) per GCU. Since each GCU is equipped with 2 GB of DDR3 memory, the available buffer is comfortably sufficient to store the full burst data, with a large safety margin.

Each GCU is equipped with 2 GB of DDR3 memory, providing a very large safety margin. In addition, the measured DDR write speed exceeds 6 GB/s, which is orders of magnitude higher than the incoming ABC–GCU data rate of 57 MB/s. Consequently, the GCU memory does not introduce any additional bottleneck and can fully absorb the data produced during a nearby CCSN, with no losses beyond those intrinsic to the front‑end electronics.

\section{Conclusions}

JUNO is the largest liquid scintillator neutrino detector ever constructed, achieving a 1\% energy linearity and a 3\% eﬀective energy resolution at 1 MeV. Its remarkable performance relies primarily on the extensive 75\% photo-coverage provided by 17,612 20-inch photomultipliers. The deployment of an additional 25,600  3-inch photomultipliers, forming the Small Photomultiplier System (SPMT), enhances the detector’s calibration capabilities addressing non-linearities, enabling high-intensity light cross-calibration, and supporting high-rate observations in specific cases.
The SPMT readout system is built around a network of 200 underwater boxes, each housing front-end electronics that interface the photodetectors with the surface electronics. Each box supports 128 photomultipliers, providing high voltage, reading and digitizing the charge, and reconstructing the arrival time of detected photons. Two 64-channel High Voltage Splitter (HVS) boards set the high voltage for each connected photomultiplier while decoupling the signals. These signals are processed by the ABC front-end readout board, equipped with eight CATIROC ASICs, each handling 16 channels to digitize charges, measure timings, and package events. A Global Control Unit (GCU) controls the ABC and HVS boards, interfaces the underwater electronics with surface systems, and connects to the data acquisition (DAQ), slow control (DCS), and low-voltage power supply.\\
A complex firmware system ensures consistent operation of the ABC and GCU boards together with the JUNO software system. Encased within a stainless steel underwater box with underwater connectors, the electronics system has been tested to operate for 20 years in JUNO’s ultrapure water. Radioactivity screening has been conducted on all components, ensuring minimal radiogenic background in agreement with the stringent single-hit rate requirements essential for JUNO’s physics goals.\\
Each board in the SPMT electronics system meets its performance specifications, validated through comprehensive laboratory testing with configurations identical to those in JUNO. The electronics achieves low noise levels of 0.04 PE, with negligible contributions to the 3-inch PMT resolution, and a crosstalk below 0.4\% across all channels. Specific tests have confirmed the system's ability to trigger single photo-electrons, accurately measure charge and timing, and transmit data with a sufficient bandwidth of approximately 57 MB/s to handle high-rate bursts such as a core-collapse supernova at 1 kpc. These results underscore the readiness of the SPMT system to contribute to JUNO’s physics program.\\
With the electronics system now in operation, the SPMT system is supporting the experiment’s physics goals. A forthcoming publication will present its commissioning results and in-situ performances.

\section{Acknowledgment}

We are gratefully for the ongoing cooperation from the China General Nuclear Power Group. This work was supported in part by: the Chinese Academy of Sciences, the National Key R\&D Program of China, the People's Government of Guangdong Province, and the Tsung--Dao Lee Institute of Shanghai Jiao Tong University in China.
We appreciate the contributions from the Institut National de Physique Nucl\'eaire et de Physique des Particules (IN2P3) in France, the Istituto Nazionale di Fisica Nucleare (INFN) in Italy, the Fonds de la Recherche Scientifique (F.R.S.--FNRS) and the Institut Interuniversitaire des Sciences Nucl\'eaires (IISN) in Belgium, the European Structural and Investment Funds, the Czech Ministry of Education, Youth and Sports and the Charles University Research Centerin Czech Republic, the Deutsche Forschungsgemeinschaft (DFG), the Helmholtz Association, and the Cluster of Excellence PRISMA+ in Germany, the Joint Institute for Nuclear Research (JINR), the Slovak Research and Development Agency in the Slovak Republic, the MOST and MOE in Taipei, the Program Management Unit for Human Resources \& Institutional Development, Research and Innovation, Chulalongkorn University, and Suranaree University of Technology in Thailand, the Science and Technology Facilities Council (STFC) in the United Kingdom, and the University of California at Irvine and the National Science Foundation (NSF) in the United States.
We also acknowledge the computing resources provided by the Chinese Academy of Sciences, IN2P3, INFN, and JINR, which are essential for data processing and analysis within the JUNO Collaboration.

\bibliographystyle{JHEP}
\bibliography{mybibliography} 

\end{document}

%% file: Authors_2025-10-22.tex
\author[39]{C\'{e}dric Cerna\thanks{Corresponding authors. Emails: cedric.cerna@in2p3.fr (C. Cerna), hem@ihep.ac.cn (M. He), jiangxs@ihep.ac.cn (X. Jiang), jpochoa@uci.edu (J.P. Ochoa-Ricoux), fperrot@lp2ib.in2p3.fr (F. Perrot).}}
\author[8]{Miao He$^*$}
\author[8]{Xiaoshan Jiang$^*$}
\author[70,5]{Juan Pedro Ochoa-Ricoux$^*$}
\author[39]{Fr\'{e}d\'{e}ric Perrot$^*$}
\author[5]{Angel Abusleme}
\author[40]{Thomas Adam}
\author[19]{Fengpeng An}
\author[68]{Costas Andreopoulos}
\author[50]{Giuseppe Andronico}

% ...
\affil[5]{Millennium Institute for SubAtomic Physics at the High-energy Frontier (SAPHIR), ANID, Chile}
% ...
\affil[8]{Institute of High Energy Physics, Beijing, China}
% ...
\affil[19]{Sun Yat-Sen University, Guangzhou, China}
% ...
\affil[39]{Univ. Bordeaux, CNRS, LP2I, UMR 5797, F-33170 Gradignan, France}
\affil[40]{IPHC, Universit\'{e} de Strasbourg, CNRS/IN2P3, F-67037 Strasbourg, France}
% ...
\affil[50]{INFN Catania and Dipartimento di Fisica e Astronomia dell Universit\`{a} di Catania, Catania, Italy}
% ...
\affil[68]{The University of Liverpool, Department of Physics, Oliver Lodge Laboratory, Oxford Str., Liverpool L69 7ZE, UK, United Kingdom}
% ...
\affil[70]{Department of Physics and Astronomy, University of California, Irvine, California, USA}
% ...


%% file: mybibliography.bib
@article{JUNO:2015zny,
    author = "An, Fengpeng and others",
    collaboration = "JUNO",
    title = "{Neutrino Physics with JUNO}",
    eprint = "1507.05613",
    archivePrefix = "arXiv",
    primaryClass = "physics.ins-det",
    doi = "10.1088/0954-3899/43/3/030401",
    journal = "J. Phys. G",
    volume = "43",
    number = "3",
    pages = "030401",
    year = "2016"
}

@article{JUNO:2021vlw,
    author = "Abusleme, Angel and others",
    collaboration = "JUNO",
    title = "{JUNO physics and detector}",
    eprint = "2104.02565",
    archivePrefix = "arXiv",
    primaryClass = "hep-ex",
    doi = "10.1016/j.ppnp.2021.103927",
    journal = "Prog. Part. Nucl. Phys.",
    volume = "123",
    pages = "103927",
    year = "2022"
}

@article{JUNO:2024jaw,
    author = "Abusleme, Angel and others",
    collaboration = "JUNO",
    title = "{Potential to Identify the Neutrino Mass Ordering with Reactor Antineutrinos in JUNO}",
    eprint = "2405.18008",
    archivePrefix = "arXiv",
    primaryClass = "hep-ex",
    doi = "10.1088/1674-1137/ad7f3e",
    journal = "Chin. Phys. C",
    volume = "49",
    number = "3",
    pages = "123001",
    year = "2025"

}

@article{JUNO:2022mxj,
    author = "Abusleme, Angel and others",
    collaboration = "JUNO",
    title = "{Sub-percent precision measurement of neutrino oscillation parameters with JUNO}",
    eprint = "2204.13249",
    archivePrefix = "arXiv",
    primaryClass = "hep-ex",
    doi = "10.1088/1674-1137/ac8bc9",
    journal = "Chin. Phys. C",
    volume = "46",
    number = "12",
    pages = "123001",
    year = "2022"
}

@article{JUNO:2023dnp,
    author = "Abusleme, Angel and others",
    collaboration = "JUNO",
    title = "{Real-time monitoring for the next core-collapse supernova in JUNO}",
    eprint = "2309.07109",
    archivePrefix = "arXiv",
    primaryClass = "hep-ex",
    doi = "10.1088/1475-7516/2024/01/057",
    journal = "JCAP",
    volume = "01",
    pages = "057",
    year = "2024"
}

@article{JUNO:2022lpc,
    author = "Abusleme, Angel and others",
    collaboration = "JUNO",
    title = "{Prospects for detecting the diffuse supernova neutrino background with JUNO}",
    eprint = "2205.08830",
    archivePrefix = "arXiv",
    primaryClass = "hep-ex",
    doi = "10.1088/1475-7516/2022/10/033",
    journal = "JCAP",
    volume = "10",
    pages = "033",
    year = "2022"
}

@article{JUNO:2020hqc,
    author = "Abusleme, Angel and others",
    collaboration = "JUNO",
    title = "{Feasibility and physics potential of detecting $^8$B solar neutrinos at JUNO}",
    eprint = "2006.11760",
    archivePrefix = "arXiv",
    primaryClass = "hep-ex",
    doi = "10.1088/1674-1137/abd92a",
    journal = "Chin. Phys. C",
    volume = "45",
    number = "2",
    pages = "023004",
    year = "2021"
}

@article{JUNO:2023zty,
    author = "Abusleme, Angel and others",
    collaboration = "JUNO",
    title = "{JUNO sensitivity to $^{7}$Be, pep, and CNO solar neutrinos}",
    eprint = "2303.03910",
    archivePrefix = "arXiv",
    primaryClass = "hep-ex",
    doi = "10.1088/1475-7516/2023/10/022",
    journal = "JCAP",
    volume = "10",
    pages = "022",
    year = "2023"
}

@article{JUNO:2021tll,
    author = "Abusleme, Angel and others",
    collaboration = "JUNO",
    title = "{JUNO sensitivity to low energy atmospheric neutrino spectra}",
    eprint = "2103.09908",
    archivePrefix = "arXiv",
    primaryClass = "hep-ex",
    doi = "10.1140/epjc/s10052-021-09565-z",
    journal = "Eur. Phys. J. C",
    volume = "81",
    pages = "10",
    year = "2021"
}

@article{JUNO:2025sfc,
    author = "Li, Yu-Feng and others",
    collaboration = "JUNO",
    title = "{Prospects for geoneutrino detection with JUNO}",
    eprint = "2511.07227",
    archivePrefix = "arXiv",
    primaryClass = "hep-ex",
    doi = "10.1088/1674-1137/ae457d",
    journal = "Chin. Phys. C in press",
    year = "2026"
}

@article{JUNO:2023ete,
    author = "Abusleme, Angel and others",
    collaboration = "JUNO",
    title = "{The Design and Technology Development of the JUNO Central Detector}",
    eprint = "2311.17314",
    archivePrefix = "arXiv",
    primaryClass = "physics.ins-det",
    doi = "10.1140/epjp/s13360-024-05830-8",
    journal = "Eur. Phys. J. Plus",
    volume = "139",
    pages = "1128",
    year = "2024"
}

@article{JUNO:2020bcl,
    author = "Abusleme, A. and others",
    collaboration = "JUNO, Daya Bay",
    title = "{Optimization of the JUNO liquid scintillator composition using a Daya Bay antineutrino detector}",
    eprint = "2007.00314",
    archivePrefix = "arXiv",
    primaryClass = "physics.ins-det",
    doi = "10.1016/j.nima.2020.164823",
    journal = "Nucl. Instrum. Meth. A",
    volume = "988",
    pages = "164823",
    year = "2021"
}

@article{JUNO:2024fdc,
    author = "Abusleme, Angel and others",
    collaboration = "JUNO",
    title = "{Prediction of Energy Resolution in the JUNO Experiment}",
    eprint = "2405.17860",
    archivePrefix = "arXiv",
    primaryClass = "hep-ex",
    doi = "10.1088/1674-1137/ad83aa",
    journal = "Chin. Phys. C",
    volume = "49",
    number = "1",
    pages = "013003",
    year = "2025"
}

@article{JUNO:2020xtj,
    author = "Abusleme, Angel and others",
    collaboration = "JUNO",
    title = "{Calibration Strategy of the JUNO Experiment}",
    eprint = "2011.06405",
    archivePrefix = "arXiv",
    primaryClass = "physics.ins-det",
    doi = "10.1007/JHEP03(2021)004",
    journal = "JHEP",
    volume = "03",
    pages = "004",
    year = "2021"
}

@article{JUNO:2023cbw,
    author = "Abusleme, Angel and others",
    collaboration = "JUNO",
    title = "{The JUNO experiment Top Tracker}",
    eprint = "2303.05172",
    archivePrefix = "arXiv",
    primaryClass = "hep-ex",
    doi = "10.1016/j.nima.2023.168680",
    journal = "Nucl. Instrum. Meth. A",
    volume = "1057",
    pages = "168680",
    year = "2023"
}

@article{Cabrera_2024,
   title={Multi-calorimetry in light-based neutrino detectors},
   volume={2024},
   ISSN={1029-8479},
   url={http://dx.doi.org/10.1007/JHEP12(2024)002},
   DOI={10.1007/jhep12(2024)002},
   number={12},
   journal={Journal of High Energy Physics},
   publisher={Springer Science and Business Media LLC},
   author={Cabrera, Anatael and others},
   year={2024},
   month=dec }

@article{JUNO:2022qgr,
    author = "Abusleme, Angel and others",
    collaboration = "JUNO",
    title = "{JUNO Sensitivity on Proton Decay $p\to \bar\nu K^+$ Searches}",
    eprint = "2212.08502",
    archivePrefix = "arXiv",
    primaryClass = "hep-ex",
    doi = "10.1088/1674-1137/ace9c6",
    journal = "Chin. Phys. C",
    volume = "47",
    number = "11",
    pages = "113002",
    year = "2023"
}

@article{Li:2019hzc,
author = {Li, Nan and Heng, Yue-kun and He, Miao and Xu, Ji-lei and Zhang, Xuan-tong and Cao, Chuan-ya and Wu, Zhi and Luo, Feng-jiao and Wang, Zhi-min and Yang, Xiao-yu and Xu, Mei-hang and Qin, Zhong-hua and Yang, Yu-zhen and Yan, Baojun and Liu, Shu-Lin},
year = {2019},
month = {03},
pages = {},
title = {Characterization of 3-inch photomultiplier tubes for the JUNO central detector},
volume = {3},
journal = {Radiation Detection Technology and Methods},
doi = {10.1007/s41605-018-0085-8}
}

@article{Xu:2025czf,
    author = "Xu, Jilei and He, Miao and Cerna, C\'{e}dric and Huang, Yongbo and others",
    title = "{Design, waterproofing, and mass production of the 3-inch PMT frontend system of JUNO}",
    eprint = "2510.06616",
    archivePrefix = "arXiv",
    primaryClass = "physics.ins-det",
    journal = "Nucl. Instrum. Meth. A",
    volume = {1086},
    pages = {171301},
    year = {2026},
    issn = {0168-9002},
    doi = {https://doi.org/10.1016/j.nima.2026.171301},
}

@article{Sato:2012jsa,
    author = "Sato, Fumitaka and Maeda, Junpei and Sumiyoshi, Takayuki and Tsukagoshi, Kento",
    editor = "Liu, Ted",
    title = "{High Voltage System for the Double Chooz Experiment}",
    doi = "10.1016/j.phpro.2012.03.735",
    journal = "Phys. Procedia",
    volume = "37",
    pages = "1164--1170",
    year = "2012"
}

@misc{axon:2024,
  author = {{Axon' Cable Ltd}},
  title = {Technical Documentation},
  howpublished = {\url{https://www.axon-interconnect.com}},
  year = {2025},
  note = {Accessed: 2026-04-27}
}

@misc{Penta:712,
  author = {{KS Plast}},
  title = {Technical Documentation},
  howpublished = {\url{https://ks-plast.ru/download/to/to_238.pdf}},
  year = {2024},
  note = {Accessed: 2024-04-27}
}

@article{JUNO:2020orn,
    author = "Conforti, Selma and others",
    collaboration = "JUNO",
    title = "{CATIROC: an integrated chip for neutrino experiments using photomultiplier tubes}",
    eprint = "2012.01565",
    archivePrefix = "arXiv",
    primaryClass = "physics.ins-det",
    doi = "10.1088/1748-0221/16/05/P05010",
    journal = "JINST",
    volume = "16",
    number = "05",
    pages = "P05010",
    year = "2021"
}

@article{SERAFINI2022167499,
    title = {The JUNO large PMT readout electronics},
    journal = "Nucl. Instrum. Meth. A",
    doi = {https://doi.org/10.1016/j.nima.2022.167499},
    volume = {1043},
    pages = {167499},
    year = {2022},
    author = {Andrea Serafini},
}

@article{WALKER2026171022,
title = {The high voltage splitter board for the JUNO SPMT system},
    journal = "Nucl. Instrum. Meth. A",
volume = {1082},
pages = {171022},
year = {2026},
issn = {0168-9002},
doi = {https://doi.org/10.1016/j.nima.2025.171022},
url = {https://www.sciencedirect.com/science/article/pii/S0168900225008241},
author = {Pablo Walker and Juan Pedro Ochoa-Ricoux and others},
}

@article{BELLATO2021164600,
title = {Embedded readout electronics R\&D for the large PMTs in the JUNO experiment},
    journal = "Nucl. Instrum. Meth. A",
volume = {985},
pages = {164600},
year = {2021},
issn = {0168-9002},
doi = {https://doi.org/10.1016/j.nima.2020.164600},
author = {M. Bellato and others}
}

@article{JUNO:2021kxb,
    author = "Abusleme, Angel and others",
    collaboration = "JUNO",
    title = "{Radioactivity control strategy for the JUNO detector}",
    eprint = "2107.03669",
    archivePrefix = "arXiv",
    primaryClass = "physics.ins-det",
    doi = "10.1007/JHEP11(2021)102",
    journal = "JHEP",
    volume = "11",
    pages = "102",
    year = "2021"
}

@article{Cao:2021wrq,
    author = "Cao, Chuanya and others",
    title = "{Mass production and characterization of 3-inch PMTs for the JUNO experiment}",
    eprint = "2102.11538",
    archivePrefix = "arXiv",
    primaryClass = "physics.ins-det",
    doi = "10.1016/j.nima.2021.165347",
    journal = "Nucl. Instrum. Meth. A",
    volume = "1005",
    pages = "165347",
    year = "2021"
}
